\documentclass[journal,twosided,web]{ieeecolor}
\usepackage{generic}
\usepackage{cite}
\usepackage{amsmath,amssymb,amsfonts}
\usepackage{caption}
\usepackage{subcaption}
\usepackage{algorithmic}
\usepackage{graphicx}
\usepackage{textcomp}
\usepackage{dsfont}
\usepackage{color}
\usepackage{epstopdf}        
\usepackage{enumerate}
\usepackage{soul}
\usepackage{lscape}
\usepackage{multicol}
\usepackage{multirow}
\usepackage{calligra}
\usepackage{booktabs}
\usepackage{mathtools}
\usepackage{framed} 
\usepackage{picins}
\usepackage{empheq}
\usepackage{afterpage}
\usepackage{graphics} 
\usepackage{epsfig} 
\usepackage{times} 
\usepackage{amsmath} 
\usepackage{amssymb}  
\usepackage{amsfonts}  
\usepackage{subfig}
\usepackage{epstopdf}
\usepackage{enumerate}
\usepackage{graphicx} 
\usepackage{xcolor}
\usepackage{comment}

\newtheorem{thm}{Theorem}
\newtheorem{prop}{Proposition}
\newtheorem{lemma}{Lemma}
\newtheorem{cor}{Corollary}
\newtheorem{definition}{Definition}
\newtheorem{assumption}{Assumption}
\newtheorem{remark}{Remark}
\newtheorem{example}{Example}
\newcommand*{\QEDB}{\hfill\ensuremath{\square}}%

\def\BibTeX{{\rm B\kern-.05em{\sc i\kern-.025em b}\kern-.08em
    T\kern-.1667em\lower.7ex\hbox{E}\kern-.125emX}}
\begin{document}
\title{Deception in Nash Equilibrium Seeking}

\author{Michael Tang, Umar Javed, Xudong Chen, Miroslav Krsti\'c, Jorge I. Poveda
\thanks{M. Tang and J. I. Poveda are with the Dep. of Electrical and Computer Engineering, University of California, San Diego, La Jolla, CA, USA.}
\thanks{U. Javed is with the Dep. of Electrical, Computer and Energy Engineering at the University of Colorado, Boulder, CO, USA.}
\thanks{ M. Krsti\'c is with the Dep. of Mechanical and Aerospace Engineering, University of California, San Diego, La Jolla, CA, USA.}
\thanks{X. Chen is with the Dep. of Electrical and Systems Engineering, Washington University in Saint Louis, MO, USA.}
\thanks{J. I. Poveda was supported in part by NSF grants ECCS CAREER 2305756, CMMI 2228791, and AFOSR YIP: FA9550-22-1-0211. Corresponding author: J. I. Poveda. {\tt jipoveda@ucsd.edu}}
}
\maketitle

\begin{abstract}
In 
socio-technical multi-agent systems, 
\emph{deception} 
exploits privileged information to induce false beliefs in ``victims,'' keeping them oblivious and leading to outcomes detrimental to them or advantageous to the deceiver.
%
We consider 
model-free {\em Nash-equilibrium-seeking} 
for non-cooperative games with asymmetric information 
and introduce  
model-free deceptive algorithms with stability guarantees. 
In the  simplest  algorithm, the deceiver includes in his action policy the victim's exploration signal, with an amplitude tuned by an integrator of the regulation error between the deceiver’s actual and desired payoff.
The integral feedback drives the deceiver's payoff to the payoff's reference value, while the victim is led to adopt a suboptimal action, at which the pseudogradient of the deceiver's  payoff is zero.
The deceiver's and victim's actions turn out to constitute a ``deceptive'' Nash equilibrium of a different game, whose structure is managed --- in real time --- by the deceiver.
{We examine quadratic, aggregative, and more general games
and provide  conditions for a successful deception, mutual and benevolent deception, and immunity to deception (for a ``non-generic'' set of payoff functions).}
Stability results are established using techniques based on averaging and singular perturbations.

Among the examples in the paper is a microeconomic duopoly in which the deceiver induces in the victim a belief that the buyers disfavor the deceiver more than they actually do, leading the victim to increase the price above the Nash price, and resulting in an increased profit for the deceiver and a decreased profit for the victim. 
A study of the deceiver's integral feedback for the desired profit reveals that, in duopolies with equal marginal costs,  a deceiver that is greedy for very high profit can attain any such profit, and pursue this with arbitrarily high integral gain (impatiently), irrespective of the market preference for the victim.
\end{abstract}

%
\vspace{-0.2cm}
\section{INTRODUCTION}
\label{sec:introduction}
\IEEEPARstart{I}{n} multi-agent systems with competitive decision-makers, the act of \emph{deception} is typically defined as the systematic exploitation of privileged information to induce false beliefs in other agents, leading them to outcomes that are detrimental to their performance or advantageous to the deceiving entity \cite{Mahon}.
%
%
%
In game theory, a player who employs deceptive strategies is termed a \emph{deceiver}. Their aim is to enhance their own outcomes covertly, often without the other players' awareness, who are typically \emph{oblivious} to such deceitful behavior. Deception in games has gained significant research attention in recent years, particularly due to concerns about safety and security in engineering systems \cite{dec_thesis}. Deception has been rigorously studied across various domains, including robotics and aerospace control \cite{dec_thesis, ho2022game}, and it 
has also been recognized as a key evolutionary feature observed in biological systems \cite{smith1987deception} and human societies \cite{mitchell1986deception}.

In this paper, we study model-free deception in N-player games, where each player aims to unilaterally minimize its own cost function. For such games, the traditional solution concept studied in the literature corresponds to the notion of a Nash equilibrium (NE) \cite{NashPaper}. When their collective actions correspond to a NE, no player has any incentive to deviate from their current action, leading to an equilibrium state for the system that is of interest in many coordination and control problems \cite{Marden09,MardenShamma,Pan:02,JCortesNash}. When the model of the cost function is \emph{unknown}, and payoff-based strategies are required, various Nash equilibrium-seeking algorithms have been studied in the literature. For example, in $N$-person non-cooperative games with individual real-valued cost functions $J_i(x)$, $i\in\{1,2,\ldots,N\}$ and vector of actions $x=[x_1,\ldots,x_N]^\top$, the work \cite{projection} introduced a class of model-free and adaptive NE seeking dynamics that implements exploration and exploitation actions in each player $i$ via the following dynamics:
\begin{subequations}\label{extsc0}
 \begin{align}
     x_i(t)&=u_i(t)+a\mu(\omega_i t)\label{extscx}\\
     \dot{u}_i(t)&=-\frac{2{k}}{a} J_i(x(t))\mu(\omega_i t),\label{extscu}
 \end{align}   
\end{subequations}
where $a,{k},\omega_i$ are positive tunable parameters, $u_i$ is an auxiliary state implemented by every player $i$, and $\mu(\omega_i t)$ is a local continuous periodic probing signal characterizing the exploration policy implemented by the agent. As shown in \cite{projection}, for a general class of games, the \emph{payoff based} algorithm \eqref{extsc0} can attain convergence to a neighborhood of the NE of the underlying game characterized by the cost functions $J_i$. Such results opened the door to the development of a variety of NE-seeking algorithms in the context of systems with delays \cite{oliveira2021nash}, games with constraints \cite{998441fb72b34a4bb2bbf3a5ba06042e}, dynamics with momentum \cite{ochoa2023momentum}, games over networks with mild coupling \cite{ditherReUse}, non-smooth algorithms \cite{PovedaKrsticBasar2020Journal}, etc. Since system \eqref{extsc0} can be seen as a continuous-time version of a discrete-time zeroth-order algorithm based on simultaneous perturbations \cite{triggeredPovedaTeel}, the results of \cite{projection} have also been connected and extended to discrete-time and stochastic settings using tools such as discrete-time averaging \cite{StankovicNashSeeking}, stochastic calculus \cite{LiuStochastic}, and stochastic approximations \cite{Nedic16,zou2022semidecentralized}, to name just a few.

Regardless of the specific NE-seeking dynamics implemented by the players, the majority of works in the literature have primarily focused on studying decision-making problems with \emph{symmetric information}, where all agents have access to the same type of signals to implement their algorithms. However, in many practical settings involving competitive and/or adversarial entities, a subset of agents has access to private information regarding the other players' algorithms. Such scenarios break the classic assumption of symmetric information, leading to games where players with privileged information can systematically use it to their advantage, sometimes called \emph{signaling games} \cite{sobel2020signaling}. How to utilize such information in NE-seeking systems without causing instabilities, while consistently improving the outcomes of players with privileged information, is a question that has not been thoroughly explored in the literature. This question motivates this work. In particular, we study a class of \emph{stable deceptive NE-seeking algorithms} that generalize \eqref{extsc0} to games with asymmetric information, where two different types of players emerge: oblivious players, who implement the classic dynamics \eqref{extsc0}, and deceptive players who implement the following policies:

\vspace{0.1cm}
\begin{subequations}\label{deceptiveNES0}
\noindent 
\emph{Zeroth-Order Exploitation Policy:}
\begin{align}   \label{commonexploitationpolicy} 
    \dot{u}_i(t)&=-\frac{2k}{a} J_i(x(t))\mu(\omega_it)
\end{align} 

\noindent 
\emph{Dynamic State-Dependent Exploration Policy:}
\begin{align} 
    x_i(t)&=u_i(t)+a\Big(\mu(\omega_i t)+ \delta_i(t) \sum_{j=1}^{n}\mu(\omega_{i_j} t)\Big),\label{decx0}\\
    \dot{\eta}_i(t)&=\varepsilon F_i\big(\eta_i(t),J_i(x(t)),u_i(t)\big),~~\varepsilon>0.\label{delta0dynamics}\\
    {\delta_i(t)}&{=h_i(\eta_i(t),x_i(t))},
\end{align}
\end{subequations}
where $F_i$ and $h_i$ are suitable smooth functions {and $\varepsilon>0$ is a small parameter designed to induce a time scale separation between the NES dynamics and the deception mechanism}.
The key difference between the nominal and deceptive NE-seeking dynamics lies in the dynamic, state-dependent exploration policy implemented by the deceivers. {In general, the deceiver can tune $\delta_i$ using only measurements of their action and their payoff.} We  show that this new policy, which  leverages privileged knowledge of the probing signals $\mu(\omega_{i_j})$ of a subset of oblivious players $\{i_j\}_{j=1}^n$, can systematically achieve closed-loop stable behaviors in the multi-agent system, while simultaneously inducing false beliefs in the oblivious players, leading them to implement actions that converge to the NE of a different game than the one being played. {This ``deceptive Nash equilibrium'' may result in better outcomes for the privileged players. As seen in \eqref{decx0}, the proposed deception mechanism takes the form of an additive modification to the deceiver's action update that incorporates the victims' exploration signals and with amplitudes updated via dynamic feedback}.


The following are the main contributions of the paper: 
%

\emph{(1)} We introduce the concept of ``deception" within the context of the NE seeking schemes proposed in \cite{projection}, which are based on algorithms that simultaneously implement exploration and exploitation policies, as in \eqref{extsc0}. We show that if a subset of the players has access to private information related to the exploration policies of other players, then they can use this information to their advantage by implementing the dynamics \eqref{deceptiveNES0} to influence the behavior of the oblivious players that implement the nominal NE seeking dynamics \eqref{extsc0}. As a result, the actions of oblivious players converge to the reaction curves of a different game, which is parameterized by the deceiving player. Specifically, we show that if a nominal, model-based, pseudogradient flow dynamic exponentially stabilizes a ``classic'' NE in a non-cooperative game, then the proposed model-free dynamics with deception will retain the exponential stability properties, but now with respect to a new \emph{Deceptive Nash Equilibrium} (DNE). We characterize the geometric properties of this DNE by studying how deception affects the reaction curves of the game learned by the oblivious players. We show that for quadratic games deception can effectively rotate or translate the reaction curves of the oblivious players, while for aggregative games deception induces nonlinear transformations that can increase the number of equilibria. By using the duopoly game as an example, we discuss various microeconomic interpretations of deception in Nash-seeking dynamics within competitive markets.
    
\emph{(2)} To align DNEs with desired individual outcomes, we show that deceptive players can modify in real-time the underlying game played by the oblivious players until the value of the cost function of the deceiver aligns with a desired reference value. This is achieved through dynamic deceptive exploration signals, managed by an auxiliary payoff-based mechanism implemented by the deceiving agents. We rigorously characterize the conditions under which the deceptive player can improve their own objective function and provide a conservative estimate of the potential improvement while maintaining the stability of the game.  We also study mutual deception when multiple deceivers have privileged information about each other's exploration policies.  
    
\emph{(3)} By leveraging the geometric characterization of the DNE, we provide sufficient conditions on the game parameters that ensure ``immunity" against deceptive players. These conditions offer valuable insights into designing counter-deception mechanisms for games and adaptive NE seeking algorithms. We also study deceptive dynamics with integral and approximate proportional action, which are able to improve transient performance to minimize potential deception detection via transient analysis. All our results are validated through simulations and numerical analysis. To the best of our knowledge, this work is the first to study stable deception mechanisms in model-free NE seeking dynamics. 

No portion of this paper was published at a conference.

Section \ref{sec_preliminaries} presents the preliminaries. Section \ref{probl_statement} introduces the problem of deceptive NE seeking. Section \ref{sec_nplayer} presents the main results for N-player quadratic games. 
Section \ref{sec_agg} focuses on deception applied to $N-$player (strongly monotone) aggregative games, and Section \ref{ref_conclusions} ends with the conclusions.

\vspace{-0.2cm}
\section{PRELIMINARIES}
\label{sec_preliminaries}
\subsubsection{Notation} We use $\langle a_{ij}\rangle$ to denote the matrix $A$ whose $(i,j)$ entry is given by $a_{ij}$. Given a matrix $Q_i$ and vector $b_i$, we use $(Q_{i})_{jk}$ to denote the $(j, k)^{th}$ entry of $Q_i$, $(Q_{i})_{j:}$ to denote the $j^{th}$ row of $Q_i$, $(Q_{i})_{:j}$ to denote the $j^{th}$ column of $Q_i$ and $(b_{i})_j$ to denote $j^{th}$ entry of $b_i$. Furthermore, we let $[Q]^{i,j}$ denote the matrix formed by taking $Q$ and removing the $i^{th}$ row and $j^{th}$ column. Having '$\sim$' in the $i$ (or $j$) entry means no row (or column) is removed. For instance, $[Q]^{\sim,3}$ is the matrix formed by only removing the third column of $Q$ (no row is removed). Given $N\in\mathbb{N}$, we use $[N]$ to denote the set of positive integers no greater than $N$, i.e $[N]:=\{1,2,...,N\}$. Given a function $f:X\to Y$ and a set $E\subset X$, we use $f(E)$ to denote the image of $f$ under $E$. A general operator $\mathcal{G}(x):\mathbb{R}^N\to\mathbb{R}^N$ is said to be $\kappa-$strongly monotone if $(\mathcal{G}(x)-\mathcal{G}(y))^\top (x-y)\ge \kappa|x-y|^2,~\forall x,y\in\mathbb{R}^N$, where $|\cdot|$ denotes the Euclidean norm and $\kappa>0$. Given $x\in\mathbb{R}$ and $r>0$, we denote the open ball of radius $r$ centered at $x$ as $B_r(x)=\{z\in\mathbb{R}: |z-x|<r\}$. We use $\mathcal{C}^k(X,Y)$ to denote the set of  $k$-times continuously differentiable functions $g:X\to Y$. Given a function $f:\mathbb{R}^n\to\mathbb{R}^m$, we let $\mathcal{N}(f)$ denote the null space of $f$, i.e $\mathcal{N}(f)=f^{-1}(\{\mathbf{0}\})$. {A function $f(x,a)$ is said to be of order $\mathcal{O}(a)$ if for each compact set $K\subset\mathbb{R}^n$ there exists $a^*,k>0$ such that $|f(x,a)|\leq ka $ for all $x\in K$ and $a\in(0,a^*)$}. 

\noindent
\subsubsection{Game Theory} We consider $N-$player non-cooperative games, where the \emph{action} of player $i$ is a scalar $x_i\in\mathbb{R}$, and each player aims to unilaterally minimize its own cost $J_i$, which also depends on the actions of the other players, and which is assumed to be continuously differentiable. We use $[N]=\{1,2,\ldots,N\}$ to denote the set of players, and $x=[x_1,...,x_N]^\top$ to denote the vector of actions. Similarly, we let $x_{-i}\in\mathbb{R}^{N-1}$ be the vector containing the actions of all players other than the $i^{th}$ player. Given real-valued cost functions $J_i(x_i,x_{-i}):\mathbb{R}^N\to\mathbb{R}$, for all $i$, a Nash Equilibrium (NE) \cite{NashPaper} is a vector $x^*\in\mathbb{R}^N$ that satisfies
\begin{equation}\label{eqNE}
x_i^*=\text{arg}\min_{x_i}J_i(x_i,x_{-i}^*),~~\forall~i\in\mathbb{R}^n,
\end{equation}
%
that is, if the other players implement $x_{-i}^*$, then no player has incentive to unilaterally deviate from $x_i^*\in\mathbb{R}$ to improve their cost. For a broad class of non-cooperative games, NE can be characterized using the \emph{pseudogradient} of the game, which is a mapping  $\mathcal{G}:\mathbb{R}^N\to\mathbb{R}^N$ given by $\mathcal{G}(x)=\left[\nabla_1 J_1(x),\dots,\nabla_N J_N(x)\right]^\top$. In particular, a NE $x^*$ satisfies $\mathcal{G}(x^*)=\mathbf{0}$, but the converse is in general not true without additional assumptions. {For each $x_{-i}\in\mathbb{R}^{N-1}$, the \emph{reaction curve} of player $i$ is defined as the set-valued mapping $\text{RC}_i(x_{-i}):\mathbb{R}^{N-1}\rightrightarrows\mathbb{R}$ that satisfies $\nabla_i J_i(\text{RC}_i(x_{-i}),x_{-i})=\{0\}$}. Since reaction curves are characterized by their graph, in this paper a ``reaction curve" will refer to the set $\mathcal{N}(\nabla_i J_i(x))$. While we assume scalar actions to avoid cumbersome notation, our models and results extend to games with vector actions via Kronecker products.
\section{STABLE DECEPTION IN GAMES WITH ASYMMETRIC INFORMATION}
\label{probl_statement}
For a general class of games with \emph{symmetric information}, 
the model-free dynamics \eqref{extsc0} have been show to attain local or semi-global practical NE seeking \cite{projection,ditherReUse,PovedaKrsticBasar2020Journal}. The key enabler for the ``learning'' capabilities of system \eqref{extsc0} are the probing signals $\mu$ that use frequencies $\omega_i\in\mathbb{R}_{>0}$, which are selected to satisfy the following assumption:
\begin{assumption}\label{assumpw}
The frequencies satisfy $\omega_i\neq \omega_j$ for $i\neq j$, and $\omega_i=\omega\bar{\omega}_i$, where $\omega\in\mathbb{R}_{>0}$ and $\bar{\omega}_i\in\mathbb{Q}_{>0}, \ \forall i\in[N]$. \QEDB 
\end{assumption}

\vspace{0.1cm}
{Assumption \ref{assumpw} is instrumental for the analysis and generalizing the set of permissible exploration frequencies beyond $\mathbb{Q}_{>0}$.} Under Assumption \ref{assumpw}, for sufficiently small values of $a$ and sufficiently large values of $\omega$, the trajectories of system \eqref{extsc0} can be approximated (on compact sets) by a perturbed pseudogradient flow of the form
\begin{equation}\label{avgsysuu}
    \dot{\tilde{u}}=-k\mathcal{G}(\tilde{u})+\mathcal{O}(a).
\end{equation}
Since nominal pseudo-gradient flows of the form $\dot{\tilde{u}}=-k\mathcal{G}(\tilde{u})$ converge to the NE in a variety of games \cite{rosen1965existence,mazumdar2020gradient}, suitable (practical) stability properties can be established for \eqref{extsc0} under an appropriate tuning of its parameters. Note that, in system \eqref{avgsysuu}, the $i^{th}$ component of $\tilde{u}$ is given by:
\begin{equation}\label{partialpseudoi}
\dot{\tilde{u}}_i=-k\nabla_iJ_i(\tilde{u})+\mathcal{O}(a),
\end{equation}
which shows that, when the $\mathcal{O}(a)$ perturbation is neglected, the equilibria of \eqref{avgsysuu} are precisely characterized by intersection of the reaction curves for all $i\in[N]$. Hence, to study deception in a general setting, we start by making the following assumption that pertains to the behavior of the unperturbed system \eqref{avgsysuu} in a nominal game with symmetric information:
\begin{assumption}\label{assumpj}
    $J_i\in\mathcal{C}^2(\mathbb{R}^N,\mathbb{R})$, there exists a unique NE $x^*\in\mathbb{R}^N$ for the game $\{J_i\}_{i=1}^N$, and the dynamics $\dot{x}=-k\mathcal{G}(x)$ render $x^*$ uniformly globally exponentially stable. \QEDB 
\end{assumption}

\vspace{0.1cm}
\begin{remark}
The stability properties of Assumption \ref{assumpj} are satisfied in a variety of games, including strongly monotone, which are common in the literature of NE seeking \cite{PovedaKrsticBasar2020Journal}. However, we note that the assumption can be relaxed to local exponential stability, which holds for a larger class of games \cite{projection}, and which can be studied via linearization methods. \QEDB 
\end{remark}

\begin{figure}
     \centering
     \begin{subfigure}[b]{0.44\textwidth}
         \centering
         \includegraphics[width=\textwidth]{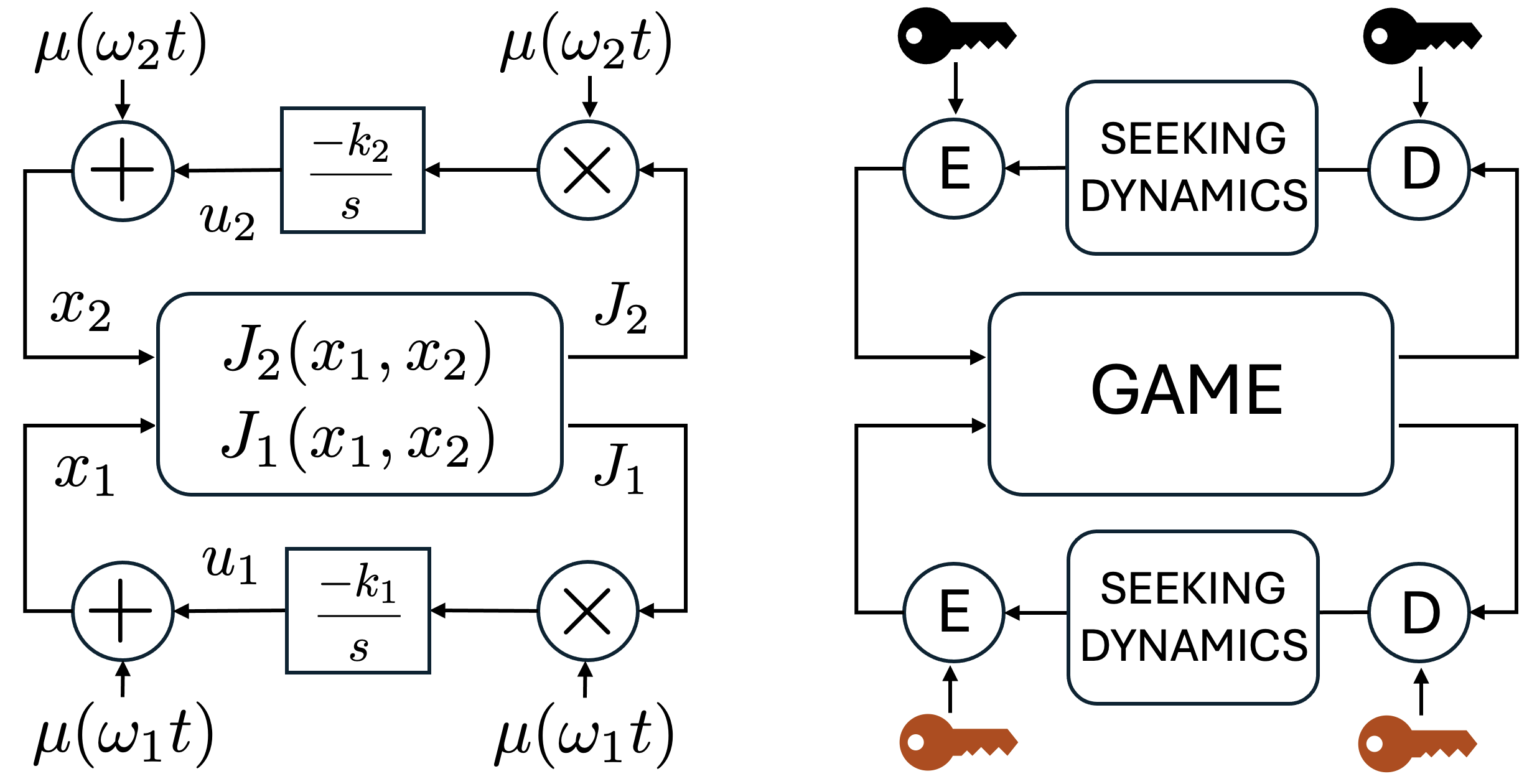}
         \label{decblockdiag}
     \end{subfigure}
     \hfill
     \begin{subfigure}[b]{0.44\textwidth}
         \centering
         \includegraphics[width=\textwidth]{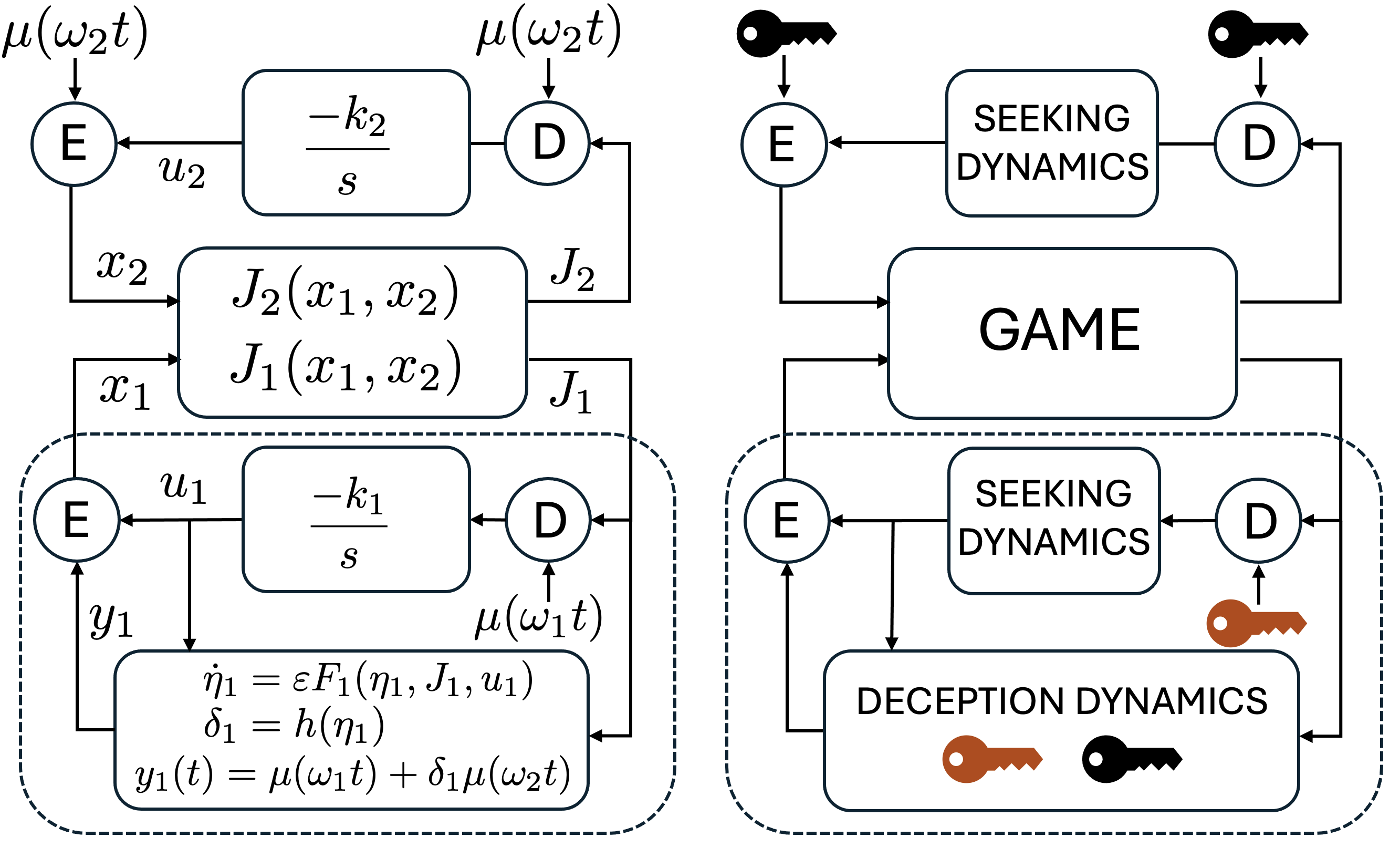}
         \label{nodecblockdiag}
     \end{subfigure}
     \hfill
        \caption{\small {Detailed and abstract block diagram representations of model-free NE seeking dynamics in 2-player games based on encryption (E) and decryption (D) using private exploration policies. Top: Scheme proposed in \cite{projection} for games with symmetric information. Bottom: Schemes studied in this paper for games with asymmetric information and deceptive players. For illustration purposes, we set $a=1$.}}
        \label{Blockdiagram1}
\end{figure}

\vspace{0.1cm}

\vspace{-0.2cm}
\subsection{Connections to Encryption/Decryption Schemes}
The principles behind systems of the form \eqref{extsc0} or \eqref{deceptiveNES0}, which simultaneously implement exploration and exploitation mechanisms, apply to other algorithms, including discrete-time and stochastic algorithms where players use random perturbations with zero expectation and identity correlation instead of periodic deterministic dither signals (see \cite{PovedaACC15, Nedic16, KusherNash}). In particular, many multi-agent model-free equilibrium-seeking algorithms rely on two key principles: 1) modulation: each agent uses an exploration signal with a unique \emph{key feature} to perturb their nominal action $u_i$; and 2) demodulation: each agent uses the same exploration signal to extract information from their cost function about their cost derivatives. This information is then used by a pre-selected exploitation policy to update $u_i$ to ``seek'' for the NE.

The two principles mentioned above happen to be identical to those employed in \emph{symmetric-key} algorithms used in cryptography for encryption and decryption of messages transmitted via communication channels in adversarial environments \cite[Ch. 2]{delfs2002introduction}. Indeed, the stability properties of a large class of algorithms based on simultaneous perturbations usually rely on conditions similar to those in Assumption \ref{assumpw}, where each player uses a different ``key" to encode and decode information about their own current assessment of the game. The left plot of Figure \ref{Blockdiagram1} illustrates this connection. This approach is reasonable in \emph{symmetric games}. However, as shown in the right plot of Figure \ref{Blockdiagram1}, in \emph{asymmetric games} where some players access others' ``keys", the privileged player's new information opens the door to novel strategic adversarial behaviors and, specially, \emph{feedback schemes} that can exploit this information and which have not been studied in NE seeking problems.

\vspace{-0.3cm}
\subsection{Nash Seeking with Deceptive and Oblivious Players}
We consider non-cooperative games with \emph{asymmetric information}, where some of the players have access to private information about the \emph{exploration policy} of other players' algorithms, and use this information to their advantage. We call such players ``deceptive''.  

\vspace{0.1cm}
\begin{definition}
A player $d\in[N]$ is said to be $\emph{deceptive}$ towards a set of players $\mathcal{D}_d:=\{d_1, d_2, ..., d_{n}\}\subset [N]$ if its actions are updated via the following rule:
\begin{subequations}\label{deceptiveNES}
\begin{align} 
    x_d(t)&=u_d(t)+a\left(\mu(\omega_d t)+ \delta_d(t) \sum_{i=1}^{n}\mu(\omega_{d_i} t)\right)\label{decx}\\ 
    \dot{u}_d(t)&=-\dfrac{2k}{a} J_d(x(t))\mu(\omega_dt),
\end{align} 
\end{subequations}
for all $t\in\text{dom}(u_d)$, where $\delta$ is updated via the \emph{deceptive dynamics}:
\begin{equation}\label{deltadynamicsgeneral}
\dot{\eta}_d(t)=\varepsilon F_d(\eta_d(t),J_d(x(t)),u_d(t)),~~\delta_d(t)=h(\eta_d(t),x_d(t)),
\end{equation}
where $\eta_d\in\mathbb{R}^m$ is an auxiliary state, $\varepsilon>0$ is a parameter, and $F_d,h$ are suitable smooth functions.  \hfill \QEDB 
\end{definition}
%
    %
%

\vspace{0.1cm}
{To simplify notation, we remove the time dependency from the states of the system}. When a player $d$ is being deceptive towards some player $d_j$, we say that player $d_j$ is \emph{oblivious} under player $d$. The set of all oblivious players under player $d$ is denoted as $\mathcal{D}_d\subset[N]$, and the set of all oblivious players in the game is denoted as $\mathcal{D}=\cup_{d\in[N]}\mathcal{D}_d\subset[N]$. Similarly, the set of all deceptive players is denoted as $\mathcal{D}^*\subset[N]$. 
%

%

\vspace{0.1cm}
In the deceptive strategy \eqref{deceptiveNES}, the parameter $\delta$ plays a crucial role. It will be used by the deceptive player to control the game that is learned in real time by the oblivious player.

%

\vspace{0.1cm}
\begin{definition}\label{defdeceptivegames}
Given a non-cooperative game $\{J_i\}_{i\in[N]}$, we say that $\{\tilde{J}_i\}_{i\in[N]}$ is a deceptive game if there exists a nonempty subset  $\mathcal{D}\subset[N]$ such that for each $i\in \mathcal{D}$ there exists $\sigma_i\in\mathcal{C}^0(\mathbb{R}^{N-1},\mathbb{R})$, a nonempty set $\mathcal{K}_i\subset[N]\setminus\{i\}$ and scalars $\delta_{j}\neq 0,~\forall j\in\mathcal{K}_i$, such that $\tilde{J}_i$ satisfies:
%
\begin{equation}\label{deceptivecost}
    \tilde{J}_i(x)=J_i(x)+\sigma_i(x_{-i})+\sum_{j\in\mathcal{K}_i}\delta_{j}\int_{0}^{x_i}\nabla_{j}  J_{i} (y)dy_i,
\end{equation}
for all $x\in\mathbb{R}^N$. If $i\not\in \mathcal{D}$, then $\tilde{J}_i(x)=J_i(x)$. \QEDB 
\end{definition}

\vspace{0.1cm}
In other words, Definition \ref{defdeceptivegames} introduces families of games $\{\tilde{J}_i\}_{i\in[N]}$ for which only the oblivious players, characterized by the index set $\mathcal{D}$, have cost functions of the form \eqref{deceptivecost} instead of the nominal cost $J_i$. Such functions depend also on the derivatives of the nominal costs $J_{i}$ with respect to the actions of the players $k\in\mathcal{K}_i$, i.e., the ``externalities'' that those players have on player $i$. 
{The $\sigma_i(x_{-i})$ term represents any $\mathcal{C}^0$ function that depends on the actions of any set of players excluding that of player $i$, thus it has no effect on the pseudogradient.} Note that, in general, both oblivious and non-oblivious players could be part of the sets $\mathcal{K}_i$ for some $i\in[N]$. However, Definition \ref{defdeceptivegames} rules out players who are ``self-deceiving'', although it leaves open the door to ``mutual deception'', which we will study in Section \ref{sec_mut}.

%

The following proposition computes the average dynamics of \eqref{extsc0} and \eqref{deceptiveNES}. The proof follows directly from the proof of Theorem \ref{thmstability} in Section \ref{sec_generalgames}.

\vspace{0.1cm}
\begin{prop}\label{prop1}
Suppose that Assumption \ref{assumpw} holds, and consider an N-player non-cooperative game $\{J_i\}_{i\in[N]}$, where a non-empty set $\mathcal{D}^*\subset[N]$ of deceptive players implements \eqref{deceptiveNES}, and the rest of the players implement \eqref{extsc0}. Then, the average dynamics of the players are given by
\begin{align}\label{averagedeceiving}
\dot{\tilde{u}}_i&=-k\nabla_i \tilde{J}_i(\tilde{u})+\mathcal{O}(a),~~~~~~~\forall~i\in[N],
\end{align}
where
\begin{align*}
\nabla_i \tilde{J}_i(\tilde{u})=\left\{\begin{array}{ll}
\hspace{-0.2cm}\nabla_i  J_i(\tilde{u})+\sum_{j\in\mathcal{K}_i}\delta_{j}\nabla_{j} 
 J_{i} (\tilde{u})&\text{if}~i\in \mathcal{D},\\
\hspace{-0.2cm}\nabla_i J_i(\tilde{u}) &\text{if}~i\not\in \mathcal{D}.
\end{array}\right.
\end{align*}
%
\QEDB
\end{prop}

The role of Proposition \ref{prop1} is to characterize the effect of the exploratory policy \eqref{decx} on the average dynamics of the Nash-seeking system. That is, \emph{deceptive} players modify the average dynamics of the \emph{oblivious players} by injecting externalities into the vector fields, each externality being weighted by the controlled parameter $\delta_{j}$. 
By effectively inducing false beliefs (i.e., $\nabla_i \tilde{J}$ instead of $\nabla_i J_i$) in the oblivious players' dynamics, deceptive players create a new pseudogradient flow \eqref{averagedeceiving} defined over the deceptive game ${\tilde{J}_i}$ instead of the nominal game ${J_i}$. If a deceptive player is not oblivious (i.e., no other player has access to his exploration policy), then his average dynamics remain unchanged and still approximate  \eqref{partialpseudoi}. In other words, deceptive non-oblivious players still learn their correct reaction curves, but they are able to control the reaction curves learned by the other players. 
\subsection{Deceptive Nash Equilibria: An Illustrative Example}
Given a deceptive N-player game $\{\tilde{J}_i\}_{i\in[N]}$, we say that $x_\delta \in \mathbb{R}^N$ is a \emph{Deceptive Nash Equilibrium} (DNE) if it is an exponentially stable equilibrium point of \eqref{averagedeceiving} with $\mathcal{O}(a)=0$ for all $i\in[N]$. In general, a DNE might not be a true NE unless it also satisfies second order conditions \cite{6736623} with respect to the deceptive game, i.e., with $\tilde{J}_i$ instead of $J_i$. Nevertheless, whether $x_{\delta}$ is a true NE of the deceptive game is unimportant in the context of stable deception. Yet, in many cases, the DNE turns out to be a NE of the deceptive game. The following example illustrates these ideas using the well-known duopoly game, extensively studied in symmetric-information games \cite{projection,PovedaKrsticBasar2020Journal,nicholson2005microeconomic}.

\vspace{0.1cm}
\begin{example}[\textbf{The Duopoly Game with Deceptive Entities}]\label{example1}
Consider a duopoly market where the two players represent companies that set the prices of their products. Let $x_1(t)$ denote the price of company 1's product at time $t$, and let $x_2(t)$ denote the price of company 2's product at the same time $t$. As shown in \cite{projection}, the negative of the profits of the companies in the market are given by
\begin{equation}\label{profitmodel}
J_i(x)=-s_i(x)(x_i-m_i),~~
\end{equation}
where, for each company $i$, $s_i$ is the number of sales, $m_i$ is the marginal cost, and $x_i-m_i$ is the profit per unit. Using the model of \cite{projection}, and assuming the consumers have a preference for the product of the company 1, as long as its price is not too large compared to the price of the product of company 2, we can model the sales using the functions $s_1(x)=S_d-s_2(x)$ and $s_2(x)=\frac{1}{p}\left(x_1-x_2\right)$, where $p>0$ quantifies the preference for company 1, and $S_d$ is the total consumer demand, assumed to be constant for simplicity. 
%
%
As shown in \cite[Sec. II]{projection}, the unique NE of this game is $x_1^*=\frac{1}{3}(2m_1+m_2+2S_dp)$, $x_2^*=\frac{1}{3}(m_1+2m_2+S_dp)$.
%
%
%
In \cite[Thm. 1]{projection}, the above game was studied under a symmetric-information assumption, and it was shown that if both players implement the NE seeking dynamics \eqref{extsc0}, then the prices $x_i$ will converge to a neighborhood of $x^*$, provided the parameters $a,\omega$ are appropriately tuned.

\begin{figure*}[t!]
  \centering
    \includegraphics[width=0.35\textwidth]{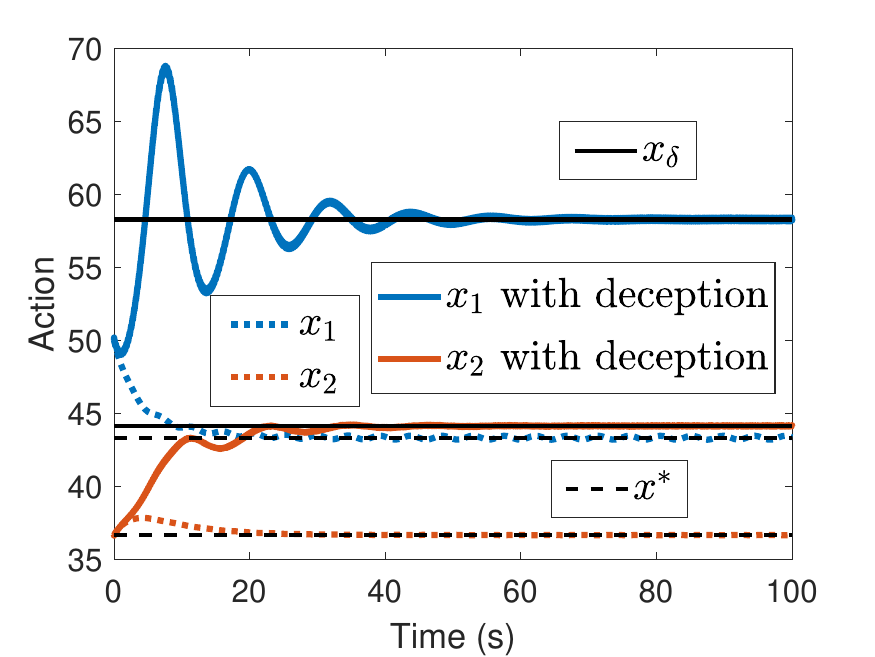}\hspace{-0.44cm}\includegraphics[width=0.35\textwidth]{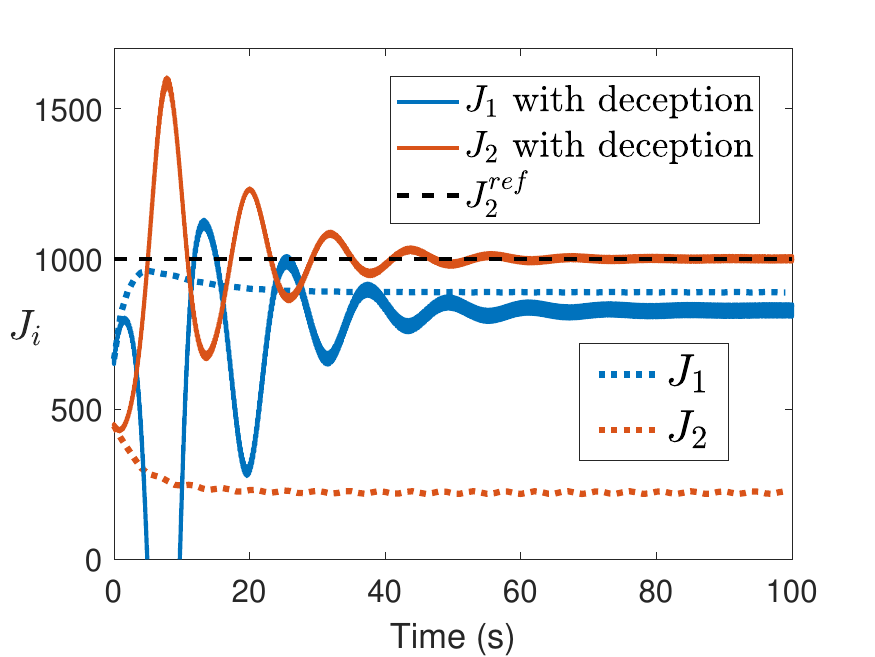}\hspace{-0.22cm}\includegraphics[width=0.32\textwidth]{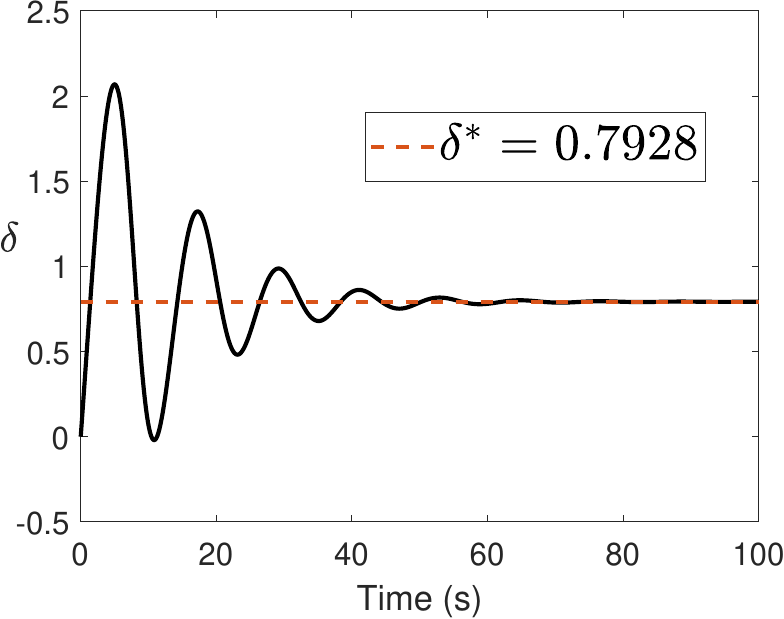}
    \caption{\small Left: A comparison for how the prices set by players 1 and 2 and the profit for player 2 changes when the deception mechanism is implemented. Center: An illustration of the payoff for player 2 with and without deception, along with the dynamics of $\delta$ (in pink) when player 2 tries to achieve $J_2^{\text{ref}}=1000$. Right: An illustration for how the RC of player 1 gets rotated in the duopoly when player 2 is deceptive. For these plots, we used $x^*=[50, 110/3]^\top, a=0.05, k=0.03, \omega_1=7877.75, \omega_2=7436.5, \varepsilon=-0.001$}\label{figduoprices}  
    \vspace{-0.2cm}
\end{figure*}

Now, suppose that company 2 has obtained knowledge of the exploration policy $\mu(\omega_1 t)$ used by player $1$, thus breaking the symmetric-information assumption. Using this knowledge, company 2 revises its strategy and implements \eqref{deceptiveNES} with a sinusoidal probing function:
%
\begin{align} \label{duox}
    x_2(t)=u_2(t)+a\Big(\sin(\omega_2 t)+\delta_2 \sin(\omega_1 t)\Big).
\end{align}
%
Company 1 is unaware of this change, and continues implementing the vanilla NE-seeking dynamics \eqref{extsc0}. The resulting average dynamics of the companies are given by
\begin{subequations}
\begin{align}\label{duoavg}
\dot{\tilde{u}}_1&=-k\Big(\nabla_1 J_1(\tilde{u})+\delta_2\nabla_2 J_1(\tilde{u})\Big)+\mathcal{O}(a),\\
\dot{\tilde{u}}_2&=-k\nabla_2 J_2(\tilde{u})+\mathcal{O}(a).
\end{align}
\end{subequations}
Neglecting the $\mathcal{O}(a)$ perturbation, this dynamics correspond to the pseudo-gradient flow of a game with costs $(\tilde{J}_1,J_2)$, where $J_2$ is still given by \eqref{profitmodel}, but $\tilde{J}_1$ is now given by \eqref{deceptivecost}, i.e., 
\begin{equation}\label{tildej1}
\tilde{J}_1(x)=J_1(x)+\sigma_1
(x_2)+\delta_2\int_{0}^{x_1}\frac{\partial J_1(y,x_2)}{\partial x_2}dy.
\end{equation}
Since $\frac{\partial J_1(y,x_2)}{\partial x_2}=-\frac{1}{p}(y-m_1)$, we have 
$\int_{0}^{x_1} \frac{\partial J_1(y, x_2)}{\partial x_2}dy=-\frac{x_1}{2p}\left(x_1-2m_1\right)$.
Thus, using \eqref{profitmodel}, the structure of $s_1$, and {choosing} $\sigma_1(x_2)=\frac{\delta_2 m_1^2}{2p}$ we have that one possible expression for $\tilde{J}_1$ is:
\begin{align}
\tilde{J}_1(x)
&=-\underbrace{\left(s_1(x)+\frac{\delta_2}{2p}(x_1-m_1)\right)}_{\text{\normalsize $=:\tilde{s}_1(x)$}}\left(x_1-m_1\right),\label{fakecost}
\end{align}
%
which, compared to \eqref{profitmodel}, now has a $\delta_2$-inflated sales function $\tilde{s}_1(x)$. Thus, whenever the price $x_1$ is above the marginal cost $m_1$, player 1 now has an incentive to increase his price further in order to increase his payoff (i.e., decrease his cost).
%

\vspace{0.1cm}
To compare the trajectories generated by both algorithms (with and without deception), we simulate the system using $S_d=100,~p=0.2,~m_1=m_2=30$, and the exploration policies $\mu(\omega_i t)=\sin(\omega_i t)$, for $i\in\{1,2\}$. To control $\delta_2$, the company 2 makes use of the following payoff-based deceptive dynamics \eqref{deltadynamicsgeneral} based on a simple integrator with state $\eta_2=\delta_2$:
\begin{equation} \label{duodelta}
    \dot{\delta}_2(t)=\varepsilon (J_2(x(t))-J_2^{\text{ref}}),
\end{equation}
where $J_2^{\text{ref}}$ is a desired profit value for the company. The left plot in Figure \ref{figduoprices} shows the trajectories $x_i$  with (solid) and without (dashed) deception. 
 As expected, when all companies implement the nominal NE-seeking dynamics \eqref{extsc0}, the prices $x$ converge to a neighborhood of the NE $x^*$. The resulting profit functions are also shown in the center plot, converging to $J_i(x^*)$, for $i \in {1, 2}$. However, note that when company 2 is deceptive, the system's trajectories converge to a neighborhood of a different action $x_\delta \in \mathbb{R}^2$, namely the DNE of the deceptive game $(\tilde{J_1},J_2)$, which, additionally, satisfies $J_2(x_\delta) = J^{\text{ref}}$, and which is given by
 \begin{subequations}
 \begin{align}
 (x_{\delta})_1&=\frac{1}{3-2\delta_2}\left((2-2\delta_2) m_1+m_2+2S_d p\right)\\
 (x_{\delta})_2&=\frac{1}{3-2\delta_2}\left((1-\delta_2)m_1 +(2-\delta_2)m_2+S_d p\right).
 \end{align}
 \end{subequations}
 The profit $J^{\text{ref}}$ is attained by company 2 using $\delta_2=\delta^*=0.7929$, shown in the right plot of Figure \ref{figduoprices}. In this way, by controlling $\delta_2$, company 2 is able to induce a different NE to attain its desired profits. In particular, company 2 deceives company 1 into overpricing relative to his Nash price, resulting in extra profit for the deceiver and reduced profit for company 1. The deception consists in making the oblivious player (i.e., company 1) believe that his sales are higher than they actually are, 
 a belief that is reflected in \eqref{fakecost}. As we will show in the next sections (see Remark 4), this type of stable deception is possible for any $J_2^\text{ref}>0$. The oscillatory transient behavior in the figure can be avoided with phase-lead compensation, as shown in Section \ref{sec_plc}.
 Note that the dynamics \eqref{extsc0}, \eqref{deceptiveNES}, and \eqref{duodelta} are all payoff-based and do not require knowledge of the mathematical form of $J_i$. Moreover, in this case, it can be verified that $x_\delta$ is also a \emph{true} NE of the deceptive game $(\tilde{J}_1, J_2)$ when $\delta_2=0.7929$. \QEDB 
\end{example}

\vspace{0.1cm}
\begin{remark}
An alternative interpretation of the deceptive mechanism can be formulated based on \emph{adaptive incentives and tolls} \cite{PhilipCSM,pricing_congestion}. In this interpretation, we write the deceptive profit as $\tilde{J}_1(x(t))=J_1(x(t))+\theta(t)$, where $\theta(t)=\delta_2(t)\int_{0}^{x_1} \frac{\partial J_1(y,x_2)}{\partial x_2}dy$ is a state-dependent \emph{incentive (or tax)}. 
Company 2 uses the deceptive NE-seeking dynamics to ``indirectly incentivize'' company 1 to choose a different price that results in higher profits for company 2. While adaptive incentives have been studied in economics \cite{nicholson2005microeconomic} and congestion problems \cite{PhilipCSM,PovedaCDC17_a,chen2023high,ochoa2022high}, we are not aware of model-free algorithms able to induce incentives/tolls via \emph{deception}. 
\QEDB 
\end{remark}
%

\vspace{-0.2cm}
\subsection{Main Result for General Games: Stable Deception}
\label{sec_generalgames}
We now generalize the previous discussions to $N$-player games with multiple deceptive and oblivious players. Consider an $N-$player non-cooperative game with $n\leq N$ deceptive players. Since we can always assign any reordering to the players, without loss of generality we assume that the set of deceptive players is given by $\mathcal{D}^*=\{1,...,n\}=[n]$, where each player $i\in[n]$ deceives a subset of players $\mathcal{D}_i:=\{d_{i,1},...,d_{i,n_i}\}$, where $n_i\in[N-1]$. To simplify our presentation, we focus on exploration policies that use sinusoidal functions, i.e., $\mu(\cdot)=\sin(\cdot)$. However, other continuous periodic functions with suitable 0-average and orthogonality properties can be used in the exploration policies.

The NES dynamics can be written in compact form as 
\begin{subequations}\label{decgamedyn}
     \begin{align}
     x_i&=\begin{dcases}
        u_i+a\left(\sin(\omega_i t)+\delta_i \sum_{j=1}^{n_i}\sin\left(\omega_{d_{i, j}}t\right)\right) & \text{if } i \in [n] \\
        u_i+a\sin(\omega_i t) & \text{else}
    \end{dcases}\label{decgame1}\\
    \dot{u}_i&=-\frac{2k}{a}J_i(x)\sin(\omega_i t),\label{decgame2}
 \end{align}
\end{subequations}
where, for simplicity, we omit writing the explicit time dependence of $x_i$ and $u_i$. To control the parameters $\delta_i$, we first focus on deceptive dynamics based on integral action
\begin{equation}\label{deltadyn}
\dot{\delta}_i=\varepsilon\varepsilon_i\left(J_i(x)-J_i^{\text{ref}}\right), \quad  i\in [n]\quad \varepsilon>0.
\end{equation}
To study the stability properties of \eqref{decgamedyn}-\eqref{deltadyn}, we define the \emph{deceptive game operator}:
%
%

\begin{equation}\label{gammamatrix}
    \gamma(\tilde{u}, \delta)=\mathcal{G}(\tilde{u})+\Lambda(\tilde{u})\delta,
\end{equation}
%
%
%
where $\delta=[\delta_1,...,\delta_n]^\top$ and
\begin{equation}\label{decperturb}
    \Lambda_{i,j}(\tilde{u})=\begin{dcases}
    \nabla_j J_i(\tilde{u}) & \text{if $i\in\mathcal{D}_j$}\\
    0 & \text{else}.
\end{dcases}
\end{equation}
We define the set:
\begin{align}\label{deltaset}
    \Delta&:=\Big\{\delta\in\mathbb{R}^n: \exists \text{ unique} \ u^*\in\mathbb{R}^N \text{ s.t } \gamma(u^*, \delta)=0 \text{ and }\notag\\&~~~ -kD_{\tilde{u}} \gamma(u^*, \delta) \text{ is Hurwitz}\Big\},
\end{align}
where $D_{\tilde{u}} \gamma$ is the Jacobian of $\gamma$ with respect to its first argument. In words, the set $\Delta$ characterizes the values of $\delta$ for which the negative pseudogradient of a deceptive game has a unique DNE. 
The next lemma follows directly from the implicit function theorem \cite[Thm 9.28]{rudin1976principles}, and it states an important property on how the DNE $u^*$ depends on $\delta$.
\begin{lemma}\label{gqss}
    Under Assumption \ref{assumpj}, the set $\Delta$ is nonempty and open, and there exists $g\in\mathcal{C}^1(\Delta, \mathbb{R}^N)$ such that $\gamma(g(\delta), \delta)=0,$ for all $\delta\in\Delta$. \QEDB 
\end{lemma}
%


%

\vspace{0.1cm}
Not every possible value of $J_i^{\text{ref}}$ may be attainable by a deceptive player under the dynamics \eqref{deltadyn}. To characterize the set of values that can be attained, we introduce the following definition, which makes use of the set $\Delta$ in \eqref{deltaset} and the function $g$ from Lemma \ref{gqss}.

\vspace{0.1cm}
\begin{definition}\label{jattaindef}
    A vector $J^{\text{ref}}=[J_1^{\text{ref}},...,J_n^{\text{ref}}]^\top$ is said to be \emph{attainable} if there exists $\delta^*\in\Delta$ such that:
    \begin{enumerate}[(a)]
        \item $J_i(g(\delta^*))=J_i^{\text{ref}},\quad\forall i\in [n]$.
        \item The matrix $\langle \nabla_j \xi_i(\delta^*)\rangle\in\mathbb{R}^{n\times n}$ is Hurwitz, where $\xi_i:\mathbb{R}^{n}\to\mathbb{R}$ is given by $\xi_i(\delta):=\varepsilon_i J_i(g(\delta))$.
    \end{enumerate}
    We let $\Omega\subset\mathbb{R}^n$ denote the set of all \emph{attainable} vectors $J^{\text{ref}}=[J_1^{\text{ref}},...,J_n^{\text{ref}}]^\top$.
    \QEDB
\end{definition}
{
\begin{remark}
    Attainability is in general a difficult condition to verify a priori, which is why we typically employ numerical methods to verify it. Fortunately, as we will see in Section \ref{sec_sdso}, the conditions are relatively simple to check for special cases of quadratic games.
\end{remark}}
\vspace{0.1cm}
The following general theorem is the first main result of the paper. It characterizes the stability properties of the NE seeking dynamics with deception:
\begin{thm}\label{thmstability}
    Consider the NE seeking dynamics \eqref{decgamedyn}-\eqref{deltadyn} with $J^{\text{ref}}\in\Omega$, namely, with $J^{\text{ref}}$ attainable, as defined in Definition 3. Suppose that Assumptions \ref{assumpw} and \ref{assumpj} hold. Then, there exists $\varepsilon^*$ such that for all $\varepsilon\in(0, \varepsilon^*)$ there exists $a^*$ such that for all $a\in(0, a^*)$ there exists $\omega^*$ such that for all $\omega>\omega^*$ the state $\zeta(t):=[u(t)\quad\delta(t)]^\top$ converges exponentially to a $\mathcal{O}(a+\frac{1}{\omega})$-neighborhood of a point $\zeta^*:=[u^*\quad\delta^*]^\top$, provided $|\zeta(0)-\zeta^*|$ is sufficiently small, where $u^*$ is the DNE.
\end{thm}
\vspace{0.1cm}

\textbf{Proof:} To analyze the system, let ${\mu}(t)=\frac{1}{a}(x-u)$, where $x, u$ are given in \eqref{decgamedyn}. Consider the time scale transformation $\tau=\omega t$, and denote $\tilde{\mu}(\tau)=\mu(\tau/\omega)$. With standard averaging theory for Lipschitz ODEs \cite{khalil}, we can compute the average dynamics of system \eqref{decgamedyn}, whose state we denote as $\tilde{u}\in\mathbb{R}^N$:
\begin{align}
    \frac{\partial{\tilde{u}}_i}{\partial \tau}&=\frac{1}{\omega T}\int_{0}^{T}-\frac{2k}{a}J_i(\tilde{u}+a\tilde{\mu}(\tau))\sin(\bar{\omega}_i \tau)d\tau\notag\\
    &=-\frac{2k}{a\omega T}\int_{0}^{T}\biggl(J_i(\tilde{u})+a\tilde{\mu}(\tau)^\top\nabla J_i(\tilde{u})\notag\\&~~~+\sum_{|\alpha|=2}\frac{(a\tilde{\mu}(\tau))^\alpha}{\alpha !}\left(\partial^\alpha J_i\right)(\ell)\biggl)\sin(\bar{\omega}_i \tau)d\tau\\
    &=-\frac{2k}{\omega T}\int_{0}^{T}\sin(\bar{\omega}_i \tau)\tilde{\mu}(\tau)^\top \nabla J_i(\tilde{u})d\tau+\mathcal{O}(a)\notag\\
    &=-\frac{k}{\omega}\left(\nabla_i J_i(\tilde{u})+\sum_{j\in\mathcal{K}_i}\delta_j \nabla_j J_i(\tilde{u})\right)+\mathcal{O}(a),\label{decavgsystem}
\end{align}
where $\ell$ is a point on the line segment connecting the points $\tilde{u}$ and $\tilde{u}+a\tilde{\mu}(\tau)$. The summation in the second equality uses multi-index notation, and we used Assumption \ref{assumpw} to evaluate the integrals. Let $\delta:=[\delta_1,...,\delta_n]^\top$, and note that the average dynamics in vector form in the original time scale are given by
\begin{equation}
    \dot{\tilde{u}}=-k\mathcal{G}(\tilde{u})-k\Lambda(\tilde{u}) \delta+\mathcal{O}(a),\label{decgameavgg}
\end{equation}
with $\Lambda(\tilde{u})$ given by \eqref{gammamatrix}. Averaging also the dynamics of $\delta$, we obtain the average system:
\begin{align}
        \dfrac{\partial \tilde{\delta}_i}{\partial \tau}&=\dfrac{\varepsilon}{\omega}\frac{1}{T}\int_{0}^{T}\varepsilon_i\left( J_i(\tilde{u}+a\tilde{\mu}(\tau))-J_i^{\text{ref}}\right)d\tau,\notag
\end{align}  
and using a Taylor series approximation in \eqref{thm1deltat}, we get
\begin{align}
        \dfrac{\partial \tilde{\delta}_i}{\partial \tau}&=\dfrac{\varepsilon}{\omega}\frac{1}{T}\int_{0}^{T}\varepsilon_i\biggl( J_i(\tilde{u})-J_i^{\text{ref}}+a\tilde{\mu}(\tau)^\top\nabla J_i(\tilde{\ell})\biggl)d\tau\label{thm1deltat}\\
        &=\dfrac{\varepsilon}{\omega}\varepsilon_i \left(J_i(\tilde{u})-J_i^{\text{ref}}\right)+\mathcal{O}(a)\quad i\in [n]\label{thm1delta}
\end{align}
where $\tilde{\ell}$ is a point on the line segment connecting the points $\tilde{u}$ and $\tilde{u}+a\tilde{\mu}(\tau)$.
Thus, the overall average dynamics of \eqref{decgamedyn}-\eqref{deltadyn} are
        \begin{subequations}\label{thm1avgsys}
        \begin{align}
        \dfrac{\partial \tilde{u}}{\partial \tau}&=\dfrac{1}{\omega}\left(-k\mathcal{G}(\tilde{u})-k\Lambda(\tilde{u}) \tilde{\delta}\right)+\mathcal{O}(a)\label{thm1u}\\
        \dfrac{\partial \tilde{\delta}_i}{\partial \tau}
        &=\dfrac{\varepsilon}{\omega}\varepsilon_i \left(J_i(\tilde{u})-J_i^{\text{ref}}\right)+\mathcal{O}(a)\quad i\in [n].\label{thm1delta2}
    \end{align}
    \end{subequations}
    Denoting ${J}^*(\tilde{u})=[J_1(\tilde{u}),...,J_n(\tilde{u})]^\top$, we have:
    \begin{align}
        \dfrac{\partial\tilde{\zeta}}{\partial\tau}&=\dfrac{1}{\omega}\begin{bmatrix}
            -k\gamma (\tilde{u}, \tilde{\delta})\\
            \varepsilon\text{diag}(\varepsilon_1,...,\varepsilon_n)\left({J}^*(\tilde{u})-J^{\text{ref}}\right)
        \end{bmatrix}+\mathcal{O}(a).\label{thm1z}
    \end{align}
    Using $\tau^*=\tau\varepsilon$ in \eqref{thm1z}, we get
    \begin{equation}\label{thm1spf}
        \begin{bmatrix}
            \varepsilon\dfrac{\partial \tilde{u}}{\partial \tau^*}\\
            \dfrac{\partial \tilde{\delta}}{\partial \tau^*}
        \end{bmatrix}=\dfrac{1}{\omega}\begin{bmatrix}
            -k\gamma(\tilde{u}, \tilde{\delta})\\
            \text{diag}(\varepsilon_1,...,\varepsilon_n)\left({J}^*(\tilde{u})-J^{\text{ref}}\right)
        \end{bmatrix}+\mathcal{O}(a).
    \end{equation}
    If we disregard the $\mathcal{O}(a)$ perturbation, the resulting system is in standard singular perturbation form, which, by Lemma \ref{gqss}, has a quasi steady state $\tilde{u}^*=g(\tilde{\delta})$. The reduced system is given by
    \begin{equation}
        \dfrac{\partial \tilde{\delta}}{\partial \tau^*}=\dfrac{1}{\omega}\text{diag}(\varepsilon_1,...,\varepsilon_n)\left({J}^*(g(\tilde{\delta}))-J^{\text{ref}}\right).\label{thm1red}
    \end{equation}
    Since $J^{\text{ref}}\in\Omega$, \eqref{thm1red} has an exponentially stable equilibrium $\delta^*\in\Delta$. Furthermore, we can also let $y:=\tilde{u}-g(\tilde{\delta})$ and obtain the boundary layer system from \eqref{thm1spf}:
    \begin{equation}
        \dfrac{\partial y}{\partial \tau}=\dfrac{1}{\omega}\left(-k\gamma(y+g(\tilde{\delta}), \tilde{\delta})\right).\label{blthm}
    \end{equation}
    where the origin is exponentially stable uniformly in $\tilde{\delta}\in \overline{B_{r^*}(\delta^*)}$ for some $r^*>0$. Thus, using a standard singular perturbation argument \cite[Ch.11.4]{khalil} (see the Appendix in the supplemental material for details) letting $u^*=g(\delta^*)$, we can find $\varepsilon^*>0$ such that for $\varepsilon\in(0, \varepsilon^*)$, $\begin{bmatrix}
        u^* & \delta^*
    \end{bmatrix}^\top:=\zeta^*$ is an exponentially stable equilibrium of the unperturbed system \eqref{thm1z}. By standard robustness results for systems with small additive perturbations, we can find $a^*>0$ such that for $a\in(0, a^*)$, $\tilde{\zeta}$ converges exponentially to a $\mathcal{O}(a)$-neighborhood of $\zeta^*$ provided $|\zeta(0)-\zeta^*|$ is sufficiently small. By averaging \cite[Thm 10.4]{khalil} we prove the claim for $\omega$ sufficiently large. \hfill $\blacksquare$

\vspace{0.1cm}
%
\begin{remark}
    As in standard NE seeking algorithms \cite{projection}, the DNE seeking dynamics can be enhanced by incorporating a phase $\phi_i$ into the exploration dither of every player, such that $\sin(\omega_i t)$ in \eqref{decgamedyn} becomes $\sin(\omega_i t+\phi_i)$. In this case, if player $d$ is deceiving player $i$, we let $\phi_{d,i}$ denote player $d$'s ``estimate" of $\phi_i$ (i.e, replace $\sin(d_{i,j}t)$ in \eqref{decgame1} with $\sin(d_{i,j}t+\phi_{i, d_{i,j}})$). Thus, $\nabla_j J_i(\tilde{u})$ in \eqref{decperturb} can be replaced with $\cos(\phi_{j,i}-\phi_i)\nabla_j J_i(\tilde{u})$, and with this modification all our results hold, i.e., as long as player $d$'s knowledge of $\phi_i$ isn't off by an odd multiple of $\frac{\pi}{2}$, the effect of deception persist. 
    \QEDB 
\end{remark}

\vspace{0.1cm}
While Theorem \ref{thmstability} is quite general and relies only on Assumptions 1-2, it does not characterize the set of attainable costs for deceptive players, or how this set relates to the stability-preserving set $\Delta$ in \eqref{deltaset}. We address these questions in the next sections by leveraging additional structures on some classes of common games studied in NE seeking problems.

%
\section{DECEPTIVE NASH EQUILIBRIUM-SEEKING IN\\ QUADRATIC GAMES}
\label{sec_nplayer}
In this section, we focus on games with costs of the form
\begin{equation}\label{quadcost}
    J_i(x)=\frac{1}{2}x^\top Q_i x + b_i^\top x +p_i,
\end{equation} 
with $Q_i\in\mathbb{R}^{N\times N}$ being symmetric $\forall i$, $b_i\in\mathbb{R}^N$ and $p_i\in\mathbb{R}$. In this case, the pseudogradient of the game is
\begin{subequations}\label{pseudog}
\begin{align}
    \mathcal{G}(x)&=\mathcal{Q}x+\mathcal{B},
\end{align}
where $\mathcal{Q}\in\mathbb{R}^{N\times N}$ and $\mathcal{B}\in\mathbb{R}^{N\times 1}$ have the form
\begin{align} 
    \mathcal{Q}&:=\begin{bmatrix}
    (Q_{1})_{1:}\\
    (Q_{2})_{2:}\\
    \vdots\\
    (Q_{N})_{N:}
    \end{bmatrix}, \ \ 
    \mathcal{B}:=\begin{bmatrix}
        (b_1)_1\\
        (b_2)_2\\
        \vdots\\
        (b_N)_N
    \end{bmatrix}.
\end{align}
\end{subequations}
When $-k\mathcal{Q}$ is Hurwitz with a negative diagonal, Assumption \ref{assumpj} holds and the NE of the game can be directly computed as $x^*:=-\mathcal{Q}^{-1}\mathcal{B}$. In this case, we can obtain an exact characterization of the average dynamics using the expansion $J_i (u+a\mu(t))=J_i(u)+a\mu(t)^\top \nabla J_i(u)+\frac{a^2}{2} \mu(t)^\top Q_i \mu(t)$.
%
%
In particular, note that under Assumption \ref{assumpw}, we have $\frac{1}{T}\int_{0}^{T} \frac{a^2}{2}\tilde{\mu}(\tau)^\top Q_i \tilde{\mu}(\tau)\sin(\bar{\omega}_i \tau)d\tau=0$, since $\frac{a^2}{2}\tilde{\mu}(\tau)^\top Q_i \tilde{\mu}(\tau)\sin(\bar{\omega}_i \tau)$ is the sum of terms that are odd and periodic in $\tau$, which implies that $\mathcal{O}(a)=0$ in \eqref{thm1u}.

To study deception in quadratic games, we introduce the  matrices $\overline{\mathcal{Q}}(i,j)\in\mathbb{R}^{N\times N}, \overline{\mathcal{B}}(i,j)\in\mathbb{R}^N$ given by:
\begin{subequations}\label{auxmatricesBQ}
\begin{align}
    \overline{\mathcal{Q}}_{k:}(i,j)&=\begin{dcases}
        (Q_{d_{i,j}})_{i:} & \text{if $k=d_{i,j}$}\\
        \mathbf{0}^\top & \text{else}
    \end{dcases}\\
    \overline{\mathcal{B}}_{k}(i,j)&=\begin{dcases}
        (b_{d_{i,j}})_{i} & \text{if $k=d_{i,j}$}\\
        0 & \text{else,}
    \end{dcases}
\end{align}
\end{subequations}
which allow to write the average dynamics \eqref{decgameavgg} as
\begin{align}
    \frac{d{\tilde{u}}}{d\tau}
    &=\dfrac{1}{\omega}\left(-k\left(\mathcal{Q}\tilde{u}+\mathcal{B}\right)-k\sum_{i=1}^{n}\sum_{j=1}^{n_i}\tilde{\delta}_i\left(\overline{\mathcal{Q}}(i,j)\tilde{u}+\overline{\mathcal{B}}(i,j)\right)\right)\notag\\
    &=\dfrac{1}{\omega}\left(-k\mathcal{Q}_{\tilde{\delta}} \tilde{u}-k\mathcal{B}_{\tilde{\delta}}\right)
    \label{decsysavg}
\end{align}
where
\begin{align*}   \mathcal{Q}_\delta&=\mathcal{Q}+\sum_{i=1}^{n}\sum_{j=1}^{n_i}{\delta}_i \overline{\mathcal{Q}}(i,j),~~\mathcal{B}_\delta&=\mathcal{B}+\sum_{i=1}^{n}\sum_{j=1}^{n_i}{\delta}_i \overline{\mathcal{B}}(i,j).
\end{align*}
Similarly, the average dynamics of $\delta$ can be computed as:
\begin{align*}
    \dfrac{\partial \tilde{\delta}_i}{\partial \tau}&=\frac{\varepsilon}{T}\int_{0}^{T}\varepsilon_i \left(J_i(\tilde{u}+a\tilde{\mu}(\tau))-J_i^{\text{ref}}\right)d\tau\\
    &=\varepsilon\varepsilon_i \left(J_i(\tilde{u})-J_i^{\text{ref}}+\dfrac{a^2}{2T}\int_{0}^{T}\tilde{\mu}(\tau)^\top Q_i \tilde{\mu}(\tau)d\tau\right)\\
    &=\varepsilon\varepsilon_i\left(J_i(\tilde{u})-J_i^{\text{ref}}+a^2\mathcal{P}_i(\tilde{\delta})\right),
\end{align*}
where $\mathcal{P}_i$ is now a quadratic function. Therefore, the set $\Delta$ in \eqref{deltaset} can be equivalently written as
\begin{equation*}
    \Delta=\{\delta\in\mathbb{R}^n : -k\mathcal{Q}_\delta \text{ is Hurwitz}\},
\end{equation*}
and the $\delta$-dependent equilibrium point of \eqref{decsysavg} is given by
\begin{equation*}
    x_\delta:=g(\delta)=-{\mathcal{Q}}_\delta^{-1}\mathcal{B}_\delta,
\end{equation*}
which is precisely the DNE of the deceptive game $\{\tilde{J}_i\}_{i\in[N]}$, and where $g$ comes from Lemma \ref{gqss}. Note that, while the set $\Delta$ is non-trivial to compute, Theorem \ref{thmstability} guarantees that $\Delta$ contains a neighborhood of $0$, and this neighborhood can be used to obtain a conservative estimate of $\Omega$.
\subsection{Deception as $\delta$-Rotations and $\delta$-Translations of Reaction Curves}
\begin{figure*}[t!]
  \centering
  \includegraphics[width=0.42\textwidth]{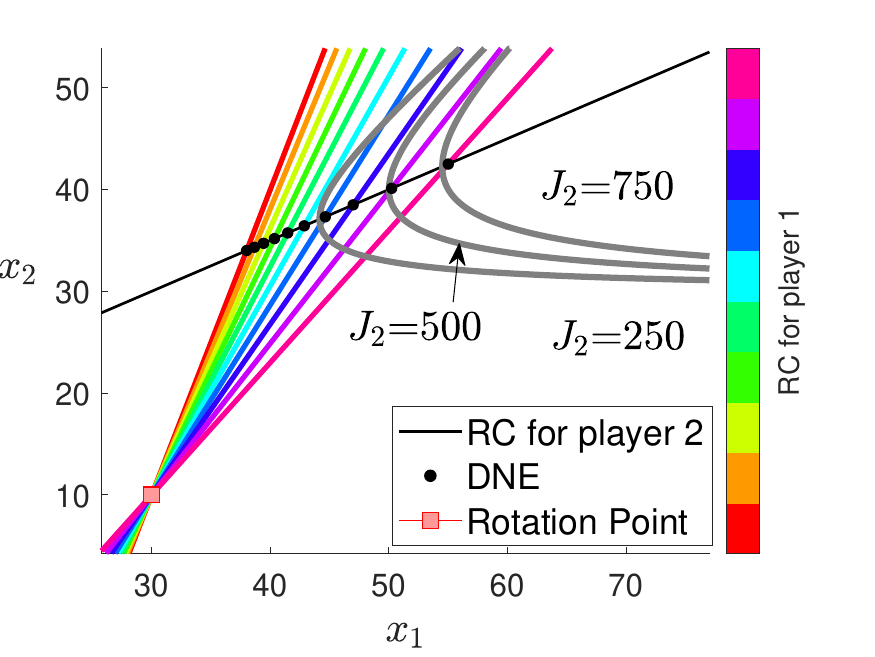}
    \includegraphics[width=0.4\textwidth]{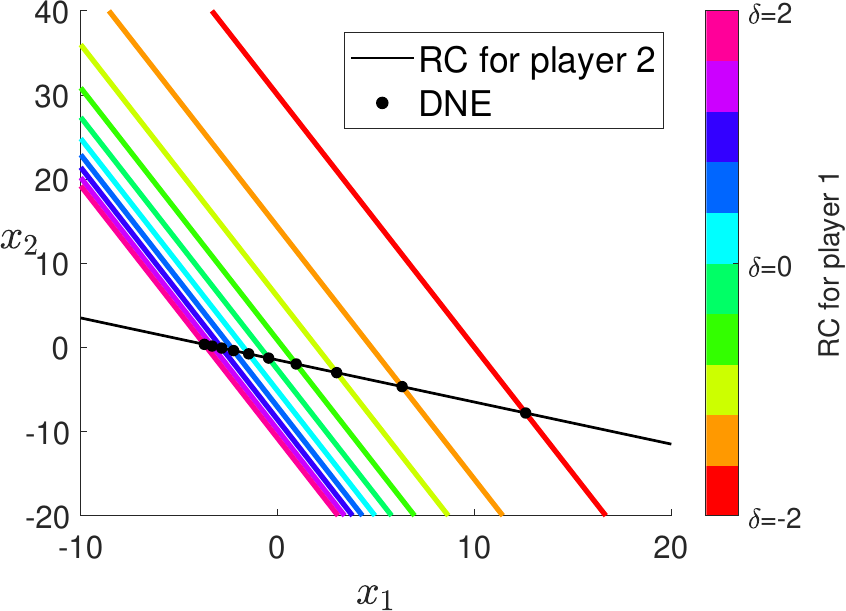}
    \caption{\small Isoprofit functions for $J_2$ and reaction curve of players under deception from player 2 in a quadratic game. Left: Duopoly game of Example 1 where deception induces rotation of RCs. Right: Quadratic game of Example 4 where deception induces translactions of RCs.}\label{rctranslation}
    \vspace{-0.2cm}
\end{figure*}
To better understand the effect of deception on quadratic games, we can study the expression in \eqref{decsysavg}. Since for $N$-player quadratic games each player's reaction curve is an $N$-dimensional affine hyperplane, it can be seen that the deception mechanism in \eqref{decgamedyn}-\eqref{deltadyn} effectively `adds' additional hyperplanes to the reaction curve of the deceived player. These hyperplanes are precisely the externalities that other players' actions have on the cost of the deceived player.

In particular, recall that the nominal reaction curve for player $k$ is given by $\mathcal{N}\left(\nabla_k J_k(x)\right)=\mathcal{N}\left((Q_k)_{k:} x+(b_k)_k\right)$. However, if player $k$ is being deceived by player $j$, the reaction curve of player $k$ in the deceptive game satisfies
\begin{align}
    \mathcal{N}&\left(\nabla_k J_k(x)+\delta_j \nabla_j J_k(x)\right)\notag\\
    &=\mathcal{N}(\left((Q_k)_{k:}+\delta_j (Q_k)_{j:}\right)x+(b_k)_k+\delta_j (b_k)_j).\label{rcdec}
\end{align}
But since the set
$\mathcal{N}(\nabla_k J_k(x))\cap \mathcal{N}(\delta_j\nabla_j J_k(x))$ does not depend on $\delta_j$ whenever $\delta_j\neq 0$, we can deduce that \eqref{rcdec} is a \emph{rotation} of player $k$'s reaction curve around the $N-1$ dimensional hyperplane 
$$
\mathcal{N}(\nabla_k J_k(x))\cap \mathcal{N}(\nabla_j J_k(x)),$$ 
provided $\mathcal{N}(\nabla_k J_k(x))\neq \mathcal{N}(\nabla_j J_k(x))$ and $\mathcal{N}(\nabla_k J_k(x))\cap \mathcal{N}(\nabla_j J_k(x))\neq \emptyset$.  On the other hand, if $\mathcal{N}(\nabla_k J_k(x))\cap \mathcal{N}(\nabla_j J_k(x))= \emptyset$, the added hyperplane is parallel to the reaction curve of player $k$, and hence \eqref{rcdec} is a \emph{translation} of player $k$'s reaction curve. The following examples illustrate these ideas.

\vspace{0.1cm}
\begin{example}[\textbf{Rotation of Reaction Curves via Deception}]
In the duopoly game of Example 1, the profits of the companies have the form \eqref{quadcost} with 
\begin{subequations}\label{duocostsj}
    \begin{align}
        Q_1&=\begin{bmatrix}
    10 & -5\\-5 & 0
\end{bmatrix},\quad b_1=\begin{bmatrix}
    -250\\150
\end{bmatrix},\quad p_1=3000\\
       Q_2&=\begin{bmatrix}
    0&-5\\-5&10
\end{bmatrix},\quad b_2=\begin{bmatrix}
    150\\-150
\end{bmatrix},\quad p_2=0
    \end{align}
\end{subequations}.
%
When company 2 implements the deceptive NE-seeking dynamics \eqref{decgamedyn}, we can use the preceding observations to verify that companies 1's \emph{perceived} reaction curve is a rotated version of the original reaction curve around the point $\mathcal{N}(\nabla_1 J_1(x))\cap \mathcal{N}(\nabla_2 J_1(x))=-Q_1^{-1}b_1=[30,10]^\top$. This rotation, for different values of $\delta_2$, is illustrated in the left plot of Figure \ref{rctranslation}, where we also show the reaction curve of company 2 (in color black) and its isoprofit functions (in color gray).\QEDB
\end{example}

\vspace{0.1cm}
\begin{example}[\textbf{Translation of Reaction Curves via Deception}]\label{exampletranslation}
Consider now a quadratic game with matrices $Q_1=[3,1;1,1/3]$, $Q_2=[1,2;2,4]$, $b_1=[7;4/3]$, and $b_2=[3;6]$, where player 2 is still deceiving and player 1 is still oblivious.
In this game, $Q_1$ is singular and $\mathcal{N}(\nabla_1 J_1(x))\cap \mathcal{N}(\nabla_2 J_2(x))=\emptyset$. Therefore, we can conclude that the ``transformation" induced on player 1's reaction curve via deception by player 2 is in fact a translation. This is visualized in the right plot of Figure \ref{rctranslation}. \QEDB 
\end{example}

\vspace{0.1cm}
{We have observed cases where deception can rotate or translate the reaction curve of the oblivious player, but it is important to note that these two phenomena cannot occur simultaneously.}
An important implication of the previous discussion is that the reaction curve of player $d_i$ is unaffected by deception if $\mathcal{N}(\nabla_k J_k(x))= \mathcal{N}(\nabla_j J_k(x))$. This property, detailed in the following lemma, characterizes a class of games that are intrinsically ``deception-immune'' under  \eqref{decgamedyn}-\eqref{deltadyn}.

\vspace{0.1cm}
\begin{lemma}[Deceptive-Immune Games]\label{decimm}
Consider an $N$-player quadratic game, and suppose that only players in $\mathcal{K}_i$ are deceptive to player $i$. If $\delta\in\Delta$, then, the following condition
\begin{equation}
         (Q_i)_{i:}=\dfrac{(b_i)_i}{(b_i)_{k}}(Q_{i})_{k :} \quad \forall k\in\mathcal{K}_i,\label{nodeccond}
\end{equation}
implies $\mathcal{N}(\nabla_i J_i(x))=\mathcal{N}(\nabla_i \tilde{J}_i(x))$. \QEDB 
\end{lemma}

\vspace{0.1cm}
\textbf{Proof:} Following the computations in \eqref{rcdec}, condition \eqref{nodeccond} implies that the reaction curve for player $i$ satisfies:
\begin{align*}
        &\mathcal{N}\left(\nabla_i J_i(x)+\sum_{k\in\mathcal{K}_i}\delta_{k} \nabla_{k} J_i(x)\right)\\
        &=\mathcal{N}\Biggl((Q_i)_{i:} x+(b_i)_i+\sum_{k\in\mathcal{K}_i}\dfrac{\delta_{k} (b_i)_{k}}{(b_i)_{i}}\left((Q_i)_{i:}x+(b_i)_{i}\right)\Biggl)\\
        &=\mathcal{N}\left(\left(1+\sum_{k\in\mathcal{K}_i}\dfrac{\delta_{k} (b_i)_{k}}{(b_i)_{i}}\right)\nabla_i J_i(x)\right)=\mathcal{N}\left(\nabla_i J_i(x)\right),
\end{align*}
where the last equality follows from the fact that $\delta\in\Delta$ implies $1+\sum_{k\in\mathcal{K}_i}\dfrac{\delta_{k} (b_i)_{k}}{(b_i)_{i}}\neq 0$. Hence, we prove the result. \hfill $\blacksquare$

\vspace{0.1cm}
Although condition \eqref{nodeccond} ``immunizes'' player $i$'s reaction curve to deception, player $i$ can still be indirectly affected if other players are deceived. However, if all deceived players' objective functions satisfy \eqref{nodeccond}, the game is immune to deception. Indeed, a sufficient condition for a fully deception-immune game is that $\text{rank}([Q_i\ | \ b_i])=1,~\forall i$. 

\vspace{0.1cm}
\begin{example}[\textbf{Immunizing Games without Changing NE}]
Consider the same quadratic game of Example 3 but with $(b_1)_2=7/3$. This new game has the same pseudo-gradient as in Example \ref{exampletranslation}, and therefore it has the same NE $x^*$. However, unlike in Example 3, the new game satisfies the assumptions of Lemma \ref{decimm}. Therefore, this game is immune to deception under the DNE-seeking dynamics \eqref{deceptiveNES}. Note that for the duopoly game studied in Example 1, the matrices $Q_1$ and $Q_2$ are invertible for any $p$, and there are no choice of parameters that can immunize the companies to deception.  \QEDB
\end{example}

\vspace{-0.2cm}
\subsection{The Single-Deceiver Single-Oblivious Player (SDSO) Case}\label{sec_sdso}
For games with one deceiver and one oblivious player, it is possible to provide a more detailed characterization of the deceiving properties of the DNE-seeking dynamics \eqref{decgamedyn}-\eqref{deltadyn}.

\vspace{0.1cm}
\subsubsection{Characterization of Attainable Costs}
In $N$-player games with one deceiver and one oblivious player, we can precisely determine how $J_i(x_\delta)$ changes with $\delta$, and whether the deceptive dynamics \eqref{deltadyn} allow deceivers to achieve any desired value $J_i^{\text{ref}}$. The following lemma is a first step in this direction:

\vspace{0.1cm}
\begin{lemma}\label{technicallemmaattainable}
    Consider a $N-$player game with quadratic costs of the form \eqref{quadcost}, and let $|\mathcal{D}^*|=|\mathcal{D}|=1$. Then
    \begin{equation*}
J_i(x_\delta)=\mathcal{J}_i(f(\delta)),~~~\forall~i\in[N],
    \end{equation*}
    where $\mathcal{J}_i$ is a quadratic polynomial and $f:\mathbb{R}\setminus\{-\frac{q_3}{q_2}\}\to\mathbb{R}\setminus\{\frac{q_1}{q_2}\}$ satisfies

    \vspace{-0.4cm}
    \begin{equation}\label{ffunction}
        f(\delta)=\dfrac{q_1 \delta}{q_2\delta+q_3},
    \end{equation}
    with $q_1,q_2,q_3\in\mathbb{R}$. If $q_3\neq0$, then $f$ is a bijection.  \QEDB 
\end{lemma}

\vspace{0.1cm}
\textbf{Proof:} We will assume without loss of generality that player 1 is deceiving player $d$, so $d_{1,1}=d$. To simplify notation we denote $\overline{\mathcal{Q}}=\overline{\mathcal{Q}}(1,1), \overline{\mathcal{B}}=\overline{\mathcal{B}}(1,1),\text{ and } \delta=\delta_1$, which means $\mathcal{Q}_\delta=\mathcal{Q}+\delta\overline{\mathcal{Q}}$ and $\mathcal{B}_\delta=\mathcal{B}+\delta\overline{\mathcal{B}}$.
Denote $e_\delta=x_\delta-x^*$. With some algebra we obtain:
\begin{equation}\label{edelteq}
    \mathcal{Q}e_\delta = \delta(-\overline{\mathcal{B}}-\overline{\mathcal{Q}}x^*-\overline{\mathcal{Q}}e_\delta).
\end{equation}
Since only the $d-$th rows of $\overline{\mathcal{Q}}$ and $\overline{\mathcal{B}}$ are nonzero, this tells us $e_\delta\in \mathcal{N}([{\mathcal{Q}}]^{d, \sim})$. If we partition $e_\delta$ as $\begin{bmatrix}
    e_{\delta,1}&e_{\delta,2:}
\end{bmatrix}^\top$, where $e_{\delta,1}\in\mathbb{R}$ and $e_{\delta,2:}\in\mathbb{R}^{N-1}$, we get:
\begin{equation}
    e_{\delta,1} ([{\mathcal{Q}}]^{d, \sim})_{:1}+[{\mathcal{Q}}]^{d,1}e_{\delta,2:}=\mathbf{0}.
\end{equation}
Solving for $e_{\delta,2:}$ and plugging into $e_\delta$ yields the following:
\begin{subequations}\label{edelta}
    \begin{align}
        e_\delta&=e_{\delta,1}\Phi\\
        \Phi&=\begin{bmatrix}1\\-\left([\mathcal{Q}]^{d,1}\right)^{-1}([{\mathcal{Q}}]^{d, \sim})_{:1}\end{bmatrix}.
    \end{align}
\end{subequations}
This formulation, of course, only makes sense if $[\mathcal{Q}]^{d,1}$ is invertible, which is another assumption we will make. Indeed, even if $[\mathcal{Q}]^{d,1}$ is singular, given that $-k\mathcal{Q}$ is Hurwitz (which implies invertibility) we can always find some $i$ such that $[\mathcal{Q}]^{d,i}$ is invertible. Then, we can select $e_{\delta,1}$ to be the $i^{th}$ entry of $e_\delta$ and partition $e_\delta$ appropriately. Without loss of generality we will make the assumption that the entry $i=1$ satisfies this property.

\par
Inspecting the $d$ row of \eqref{edelteq} to solve for $e_{\delta,1}$, we obtain
\begin{align} \label{fdelta}
    f(\delta)&:=e_{\delta,1}=\frac{q_1\delta}{q_2 \delta+q_3},
    \end{align}
where
\begin{equation*}
    q_1=-((b_d)_1+(Q_d)_{1:}x^*),~~q_2=(Q_d)_{1:}\Phi,~~q_3=(Q_d)_{d:}\Phi.
\end{equation*}
Using \eqref{fdelta} we directly obtain  $ J_{i}(x_\delta)=\frac{1}{2}e_\delta^\top Q_i e_\delta+(Q_i x^*+b_i)^\top e_\delta+J_i(x^*)$.
%
Substituting \eqref{edelta} into this expression leads to 

\begin{equation}\label{quadj}
    \mathcal{J}_i(e_{\delta,1})=r_{i,2} e_{\delta,1}^2+r_{i,1} e_{\delta,1}+J_i(x^*),
\end{equation}
where
\begin{equation}\label{rterms}
    r_{i,2}=\frac12\Phi^\top Q_i \Phi,\quad r_{i,1}=(Q_i x^*+b_i)^\top \Phi,
\end{equation}
which establishes the result. \hfill $\blacksquare$


\vspace{0.1cm}
By leveraging Lemma \ref{technicallemmaattainable}, we can now characterize the structure of the set of attainable costs $\Omega$ for SDSO quadratic games:

\vspace{0.1cm}
\begin{thm}\label{omegaval}
      Consider a $N$-player quadratic SDSO game with cost functions \eqref{quadcost}, $\mathcal{D}^*=\{1\}$, $\mathcal{D}_1=\{d\}$, and let $\mathcal{J}_1$ be given by \eqref{quadj} and $f$ be given by \eqref{ffunction}.
      \begin{enumerate}[(a)]
      \item If $\varepsilon_1 r_{1,2} q_1 q_3>0$, then
      \begin{equation*}
        \Omega=\mathcal{J}_1\left(\left(-\infty, -\frac{r_{1,1}}{2r_{1,2}}\right)\bigcap f(\Delta)\right).
      \end{equation*}
      \item If $\varepsilon_1 r_{1,2} q_1 q_3<0$, then
      \begin{equation*}
    \Omega=\mathcal{J}_1\left(\left(-\frac{r_{1,1}}{2r_{1,2}}, \infty\right)\bigcap f(\Delta)\right).
      \end{equation*}
      \end{enumerate}
\end{thm}

\vspace{0.2cm}
\textbf{Proof:} Since there is only one deceptive player, we have $\delta=\delta_1$. Using the definition of $\xi$ in Definition \ref{jattaindef}, and the result from Lemma \ref{technicallemmaattainable}, we obtain that
\begin{align*}
    \frac{\partial \xi_1}{\partial \delta}&
    =\varepsilon_1\frac{\partial \mathcal{J}_1(f(\delta))}{\partial \delta}=\frac{2\varepsilon_1 r_{1,2} q_1 q_3 \left( f(\delta)+\dfrac{r_{1,1}}{2r_{1,2}}\right)}{(q_2\delta+q_3)^2}<0
\end{align*}
where $r_{i,j}$ is defined in \eqref{rterms}. The result follows now directly from Definition \ref{jattaindef}. \hfill $\blacksquare$
\vspace{0.1cm}
\begin{remark}
    Using these results, we obtain that for the duopoly of Example 1,  $\Delta=(-\infty, 1.5)$ and $\Omega=(0, \infty)$. However, as $J_2^{\text{ref}}\to \infty$, we have $\delta^*\to 1.5$. So, as the deceiver gets greedier, the basin of attraction of the DNE shrinks. \QEDB 
\end{remark}

\vspace{0.1cm}
\begin{example}[\textbf{On the Geometry of Attainable Costs}]
Consider a 2-player quadratic game with cost parameters $Q_1=[3,1;1,5]$, $b_1=[4;2]$, $Q_2=[7,2;2,4]$, $b_2=[1;6]$. Let player 2 be a deceptive player towards player 1. In this case, we have $\mathcal{Q}=[3,1;2,4]$, $\mathcal{B}=[4;6]$, and the NE is $x^*=[-1;-1]$.  We can use our previous theoretical results to compute the key properties of the emerging DNE. Specifically, using \eqref{auxmatricesBQ} we obtain $\overline{\mathcal{Q}}=[1,5;0,0]$ and $\overline{\mathcal{B}}=[2;0]$. Similarly,
%
    $\Delta=(-7,\frac{5}{3})$, and using \eqref{edelta} we obtain $e_\delta=e_{\delta,1}\Phi$, where $\Phi=\begin{bmatrix}
    1&-0.5
\end{bmatrix}^\top$ and $e_{\delta,1}=4\delta\left(-1.5\delta+2.5\right)^{-1}$. It follows that player 2 can rotate the reaction curve of player 1 around the point $-Q_1^{-1} b_1=[-1.286;-0.143
]$. Lastly, we have $\mathcal{J}_2(e_{\delta,1})=3e_{\delta,1}^2-8e_{\delta,1}+0.5$,
which achieves a minimum of -4.83 at $e_{\delta,1}^*=\frac{4}{3}\to \delta=\frac{5}{9}$. Using Theorem \ref{omegaval} we can compute the range of attainable $J_2$ values, which is $\Omega=(-4.83, 31.64)$. \QEDB  \end{example}

In words, the result of Theorem \ref{omegaval} characterizes the level of profits that players can achieve via deception in quadratic games when there is only one deceiver and one oblivious player. Since, in practice, the parameters of the game are unknown, this characterization is only relevant to establish \emph{viability} for deception. Note that if $J^{\text{ref}}$ is not moderate, one might have $J^{\text{ref}}\notin\Omega$, and then instability can emerge, i.e., \emph{greedy deception might lead to instability}.

\vspace{0.1cm}
\subsubsection{Benevolent Deception}
In the duopoly example, it can be verified that setting $J_2^{\text{ref}}=1000$ shifted the Nash equilibrium to a position more favorable for company 2 but less favorable for company 1. However, in some cases deception can also benefit oblivious players. This setting is known in the literature as benevolent deception \cite{adar2013benevolent}, and it can also emerge under the DNE seeking dynamics \eqref{decgamedyn}-\eqref{deltadyn} with an appropriate choice of $J_i^{\text{ref}}$ in \eqref{deltadyn}. 

To study benevolent deception, note that the cost functions in the duopoly game, evaluated at the DNE, can be written as follows using \eqref{fdelta}:
\begin{subequations}
    \begin{align*}        J_1(x_\delta)&=-2.5e_{\delta,1}^2+33.3e_{\delta,1}+J_1(x^*)\\   J_2(x_\delta)&=1.25e_{\delta,1}^2+33.3e_{\delta,1}+J_2(x^*).
    \end{align*}
\end{subequations}

To improve \emph{both} payoff functions, $\delta$ must satisfy
\begin{align*}
    \delta&\in \{\delta: -2.5e_{\delta,1}^2+33.3e_{\delta,1}>0, -1.25e_{\delta,1}^2+33.3e_{\delta,1}>0\}\\&=(0, 0.75).
\end{align*} 
Which is consistent with the fact that in Example \ref{example1} the desired value $J_2^{\text{ref}}=1000$ was achieved with $\delta=0.7929\not\in(0, 0.75)$. To characterize the values of $J_2^{\text{ref}}$ that lead to benevolent deception, we can compute the set $J_2(g((0, 0.75)))=(222.2, 888.8)$, where $g(\delta)$ is the DNE (from Lemma \ref{gqss}). It follows that for any $J_2^{\text{ref}}\in(222.2, 888.8)$, the seeking dynamics \eqref{decgamedyn}-\eqref{deltadyn} will stabilize a DNE that leads to a better payoff for both players. These ideas can be generalized to an $N-$player setting. Using \eqref{quadj}, let 
\begin{equation*}
\mathcal{F}_i(e_{\delta,1}):=\mathcal{J}_i(e_{\delta,1})-J_i(x^*)=r_{i,2} e_{\delta,1}^2+r_{i,1} e_{\delta,1},
\end{equation*} 
and note that $ \mathcal{F}_i'(0)=r_{i,1}$. The following theorem establishes benevolent DNE-seeking whenever player 1 is the deceptive player and player $d$ is the oblivious player.



\vspace{0.1cm}    
\begin{thm}\label{thmbenevolent}
    Consider an N-person non-cooperative game with cost functions \eqref{quadcost} and players implementing the DNE-seeking dynamics \eqref{decgamedyn}-\eqref{deltadyn}, where $n=1$, $n_1=1$, and $d_{1,1}=d$. Let $\mathcal{M}$ denote a subset of players with cost functions satisfying
    \begin{equation*}
        \text{sgn}\left(r_{1,1}\right)=\text{sgn}\left(r_{i,1}\right),~~ \forall i\in\mathcal{M}.
    \end{equation*}
    Suppose $\varepsilon_1 r_{1,1}q_1 q_3<0$. Then $\Omega$ is nonempty and there is a nonempty subset $\Omega^*\subset\Omega$ such that for each $J_1^{\text{ref}}\in\Omega^*$, it follows that $J_i(u^*)<J_i(x^*)$ for all $i\in \mathcal{M}\cup \{1\}$, where $\zeta^*=\begin{bmatrix}
        u^*\quad \delta^*
    \end{bmatrix}^\top$ is the DNE generated by Theorem \ref{thmstability} with $J^{\text{ref}}=J_1^{\text{ref}}$. \QEDB 
\end{thm}

\textbf{Proof:} First we assume $r_{1,1}>0$, and the $r_{1,1}<0$ case is nearly identical. This means $\mathcal{F}_i'(0)>0$ for each $i\in\mathcal{M}\cup\{1\}$, so we can find some $R>0$ such that $\mathcal{F}_i(e_{\delta,1})<0$ whenever $e_{\delta,1}\in(-R, 0), \ \forall i\in\mathcal{M}\cup\{1\}$. We also have 
    \begin{align*}
        \dfrac{\partial \xi_1}{\partial \delta}\bigg |_{\delta=0}&=\dfrac{2\varepsilon_1 r_{1,2} q_1 q_3 \left( f(\delta)+\dfrac{r_{1,1}}{2r_{1,2}}\right)}{(q_2\delta+q_3)^2}\bigg |_{\delta=0}=\dfrac{\varepsilon_1 r_{1,1} q_1 q_3}{q_3^2},
    \end{align*}
    which is negative by our assumptions, so we can find $R^*>0$ such that $\dfrac{\partial \xi_1}{\partial \delta}<0$ for all $\delta\in B_{R^*}(0)\subset\Delta$. We can define the following set
    \begin{equation*}
        \mathcal{E}=(-R,0)\cap f(B_{R^*}(0)),
    \end{equation*}
    which is nonempty since $f$ is continuous and strictly monotone in a neighborhood of $0$. We can then let $\Omega^*=\mathcal{J}_1(\mathcal{E})$, which completes the proof. \hfill $\blacksquare$
    
\begin{figure}[t!]
  \centering
    \includegraphics[width=0.45\textwidth]{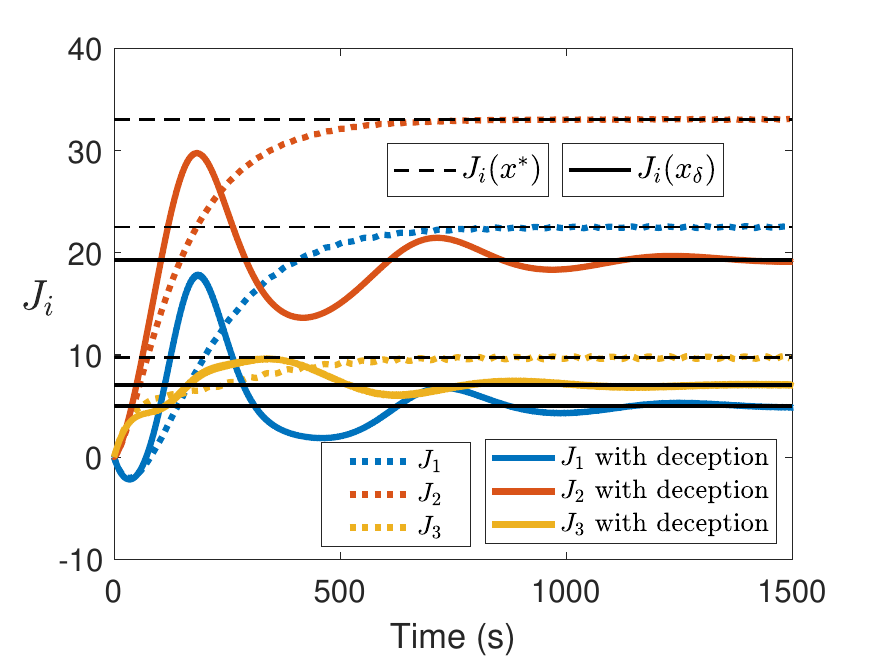}
    \caption{\small Price convergence of a 3-player quadratic game with benevolent deception, where player 1 deceives player 3 with $J_1^{\text{ref}}=-1$, 
    $Q_1=[0.7,0.25,-0.1;0.25,0.6,0.05;-0.1,0.05,0.9], b_1=[2;2;-3], Q_2=[0.7,-0.15,0.05;-0.15,0.8,-0.1;0.05,-0.1,0.2],\\ b_2=[-1;-3;3], Q_3=[-0.15,0,0.125;0,0.1,0.05;0.125,0.05,0.35],\\ b_3=[2;7;-3]$ and the players use $a=0.04, k=0.02, \omega_1=3172.8, \omega_2=2044.4, \omega_3=3057.6$}\label{3ex}
    
\end{figure}
\vspace{0.1cm}

\vspace{0.1cm}
\begin{example}[\textbf{Benevolent Deception in Multi-Player Games}]
\label{sec_numexamples}
%

%
%
%
To provide more insight on our results and show their application in larger multi-player games, we consider a 3-player quadratic game with cost functions having parameters defined in the caption of Figure \ref{3ex}. The pseudogradient parameters are given by
\begin{equation*}
    \mathcal{Q}=\begin{bmatrix}
        0.7&0.25&-0.1\\-0.15&0.8&-0.1\\
        0.125&.05&0.35
    \end{bmatrix},\quad\mathcal{B}=\begin{bmatrix}
        2\\-3\\-3
    \end{bmatrix}.
\end{equation*}
If player 1 is deceptive to player 3, we have
\begin{equation*}
    \overline{\mathcal{Q}}=\begin{bmatrix}
        0&0&0\\0&0&0\\-0.15&0&0.125
    \end{bmatrix}
    ,\quad\overline{\mathcal{B}}=\begin{bmatrix}
        0\\0\\2
    \end{bmatrix}.
\end{equation*}
An application of the Routh-Hurwitz criteria yields $\Delta=(-3.315, \infty)$, and from the proof of Lemma \ref{technicallemmaattainable} we also have $\Phi=[1;1.55;10.86]$ and $f(\delta)=-3.6\delta(1.21\delta+4)^{-1}$. We set $\varepsilon_1>0$, and with Proposition \ref{omegaval} we obtain $\Omega=\mathcal{J}_1((-0.629, \infty))=(1.04, \infty)$. Furthermore, we also have $r_{1,1}=68.3, r_{2,1}=42.6$ and $ r_{3,1}=15.8$. When $\varepsilon_1>0$, the conditions of Theorem \ref{thmbenevolent} are satisfied, which guarantees the possibility of benevolent deception. We also have $J_1(x^*)=22.5$, so if player $1$ chooses $J_1^{\text{ref}}=5$, which is achieved at $\delta=0.454$, it can be verified that each player's cost function improves when the NE is moved to the stable DNE. Moreover, it can also be verified that the DNE is a true NE of the deceptive game. The results are verified and illustrated in Figure \ref{3ex}. \QEDB 
\end{example}

\subsection{Mutual Deception}\label{sec_mut}
In this section, we study an important question that can emerge in systems with more than one deceiver and oblivious player: what happens when two players are deceptive to each other? This situation is refer to in the literature as \emph{mutual deception} \cite{lopez2012power}. {This is a special case of the generalized deception considered in Theorem \ref{thmstability}, but it raises some interesting questions.}

To study mutual deception in non-cooperative games, we focus on two-player games with exploration policies given by
\begin{equation}\label{mutualdecmu}
    \mu(t)=\begin{bmatrix}
        \sin(\omega_1 t)+\delta_1 \sin(\omega_2 t)\\
        \sin(\omega_2 t)+\delta_2 \sin(\omega_1 t)
    \end{bmatrix}.
\end{equation}
In this case, the average dynamics of the system are given by \eqref{decsysavg} with $\delta=[\delta_1\quad\delta_2]^\top$ and
\begin{subequations}\label{mutualparameters}
\begin{align}
    \mathcal{Q}_\delta&=\mathcal{Q}+\delta_1 \begin{bmatrix}
        \mathbf{0}^\top\\(Q_2)_{1:}
    \end{bmatrix}+\delta_2 \begin{bmatrix}
        (Q_1)_{2:}\\\mathbf{0}^\top
    \end{bmatrix}\\
    \mathcal{B}_\delta&=\mathcal{B}+\delta_1\begin{bmatrix}
        0\\(b_2)_1
    \end{bmatrix}+\delta_2\begin{bmatrix}
        (b_1)_2\\0
    \end{bmatrix}.
\end{align}
\end{subequations}
We let each player $i$ update $\delta_i$ via the first-order integrator dynamics:
\begin{equation*}
    \dot{\delta}_i=\varepsilon\varepsilon_i \left(J_i(x)-J_i^{\text{ref}}\right),
\end{equation*}
where $\varepsilon>0$.
In this case, by direct computation we can observe that the vector $[J_1^{\text{ref}}, J_2^{\text{ref}}]$ is attainable if there is a $\delta^*=[\delta_1^*\quad \delta_2^*]^\top$ for which the following three properties hold:
\begin{enumerate}[(a)]
    \item $-k\mathcal{Q}_{\delta^*}$ is Hurwitz
    \item $J_i(-\mathcal{Q}_{\delta^*}^{-1}\mathcal{B}_{\delta^*})=J_i^{\text{ref}}$ for $i=1,2$
    \item The following matrix is Hurwitz: $$\begin{bmatrix}
            \dfrac{\partial\xi_1}{\partial \delta_1}(\delta^*)&\dfrac{\partial\xi_1}{\partial \delta_2}(\delta^*)\\
            \dfrac{\partial\xi_2}{\partial \delta_1}(\delta^*)&\dfrac{\partial\xi_2}{\partial \delta_2}(\delta^*)
        \end{bmatrix},$$ where $\xi_i=\varepsilon_i J_i(g(\delta))=\varepsilon_i J_i(-\mathcal{Q}_{\delta}^{-1}\mathcal{B}_{\delta})$.
\end{enumerate}
%

\vspace{0.1cm}
\noindent
To illustrate these ideas, we  revisit the duopoly example when company 1 is also deceptive to company 2.

\begin{figure*}[t]
  \centering
    \includegraphics[width=0.31\textwidth]{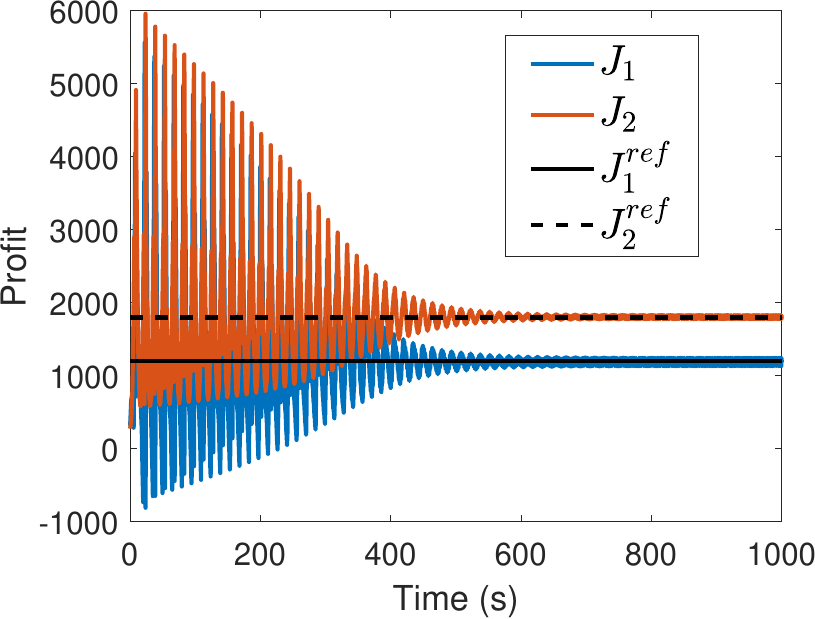}
    \hspace{-0.1cm}\includegraphics[width=0.33\textwidth]{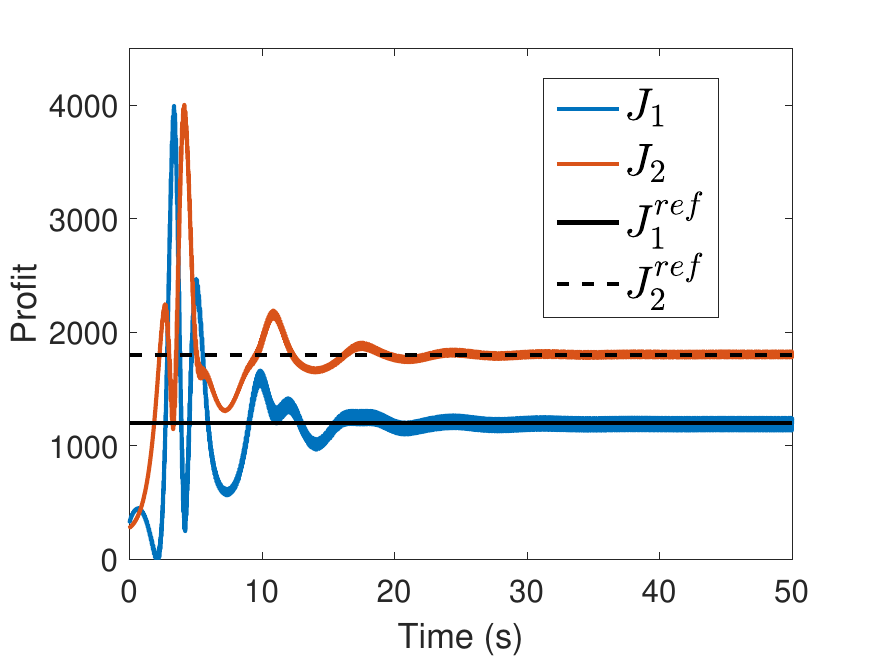}\includegraphics[width=0.34\textwidth]{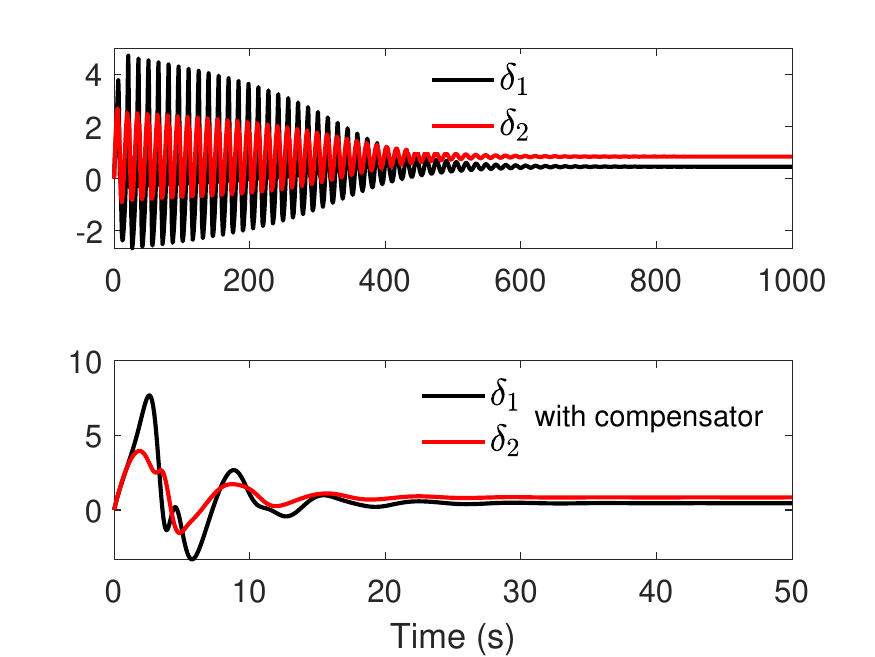}
    \hspace{-0.1cm}
    \caption{\small Left: Profit convergence in a duopoly game under mutual deception using first-order deceptive dynamics. Center: Profit convergence in a duopoly game under (mutual) deception using second-order deceptive dynamics. Right: trajectories of $\delta$ in a mutually deceptive duopoly with first and second order tuning dynamics on $\delta$}\label{figmut}
    \vspace{-0.5cm}
\end{figure*}

\vspace{0.1cm}
\begin{example}[\textbf{Duopoly Game with Mutual Deception}]
For the duopoly game studied in Example \ref{example1}, we show that the effect of deception of company 2 on company 1 was reflected on a modified cost $\tilde{J}_1$ with an inflated sales function. Using similar computations, it can be verified that if company 1 is also deceptive to company 2, the perceived cost of company 2 becomes:
\begin{align}\label{modifiedcost2}
\tilde{J}_2(x)
&=-\frac{1}{p}\left(\tilde{x}_1-\tilde{x}_2\right)(x_2-m_2)+\sigma_2(x_1),
\end{align}
where $\tilde{x}_1=x_1-\frac{\delta_1}{2}m_2$, $\tilde{x}_2=\left(1-\frac{\delta_1}{2}\right)x_2$ and $\sigma_2(x_1)=\frac{\delta_1 m_2^2}{2p}$. In \eqref{modifiedcost2}, company 2's perceived cost has now a distorted sales value $s_2$ that depends on deceiving prices $\tilde{x}_1$,~$\tilde{x}_2$. Note that the resulting quadratic game can be written as \eqref{quadcost} with
\begin{subequations}
    \begin{align*}
        \mathcal{Q}_\delta&=\begin{bmatrix}
            10&-5\\-5&10
        \end{bmatrix}+\delta_1\begin{bmatrix}
            0&0\\0&-5
        \end{bmatrix}+\delta_2\begin{bmatrix}
            -5&0\\0&0
        \end{bmatrix}\\
        \mathcal{B}_\delta&=\begin{bmatrix}
            -250\\-150
        \end{bmatrix}+\delta_1\begin{bmatrix}
            0\\150
        \end{bmatrix}+\delta_2\begin{bmatrix}
            150\\0
        \end{bmatrix}.
    \end{align*}
\end{subequations}
Suppose the companies implement \eqref{deltadyn} with $J_1^{\text{ref}}=1200, J_2^{\text{ref}}=1800$, leading to $\delta_1^*=0.459$ and $\delta_2^*=0.848$. With $\varepsilon=0.001, \varepsilon_1=1$ and $ \varepsilon_2=\frac12$, the second condition of Definition \ref{jattaindef} leads to:
\begin{equation*}
        \begin{bmatrix}
            \dfrac{\partial\xi_1}{\partial \delta_1}(\delta^*)&\dfrac{\partial\xi_1}{\partial \delta_2}(\delta^*)\\
            \dfrac{\partial\xi_2}{\partial \delta_1}(\delta^*)&\dfrac{\partial\xi_2}{\partial \delta_2}(\delta^*)
        \end{bmatrix}=\begin{bmatrix}
            -2007.3&3129.3\\
            -1011&-3577.2
        \end{bmatrix},
\end{equation*}
which is Hurwitz. It can also be verified that $-k\mathcal{Q}_{\delta^*}$ is Hurwitz. Thus, the solutions of the closed-loop system converge to the DNE. This can be visualized in the left plot of Figure \ref{figmut}, where the DNE-seeking dynamics used $a=0.05$, $k=-0.03$, $\omega_1=7877.75$, and $\omega_2=7436.5$. {This example also showcases the possibility of benevolent deception and mutual deception occurring simultaneously.}
%
\end{example}

\subsection{Adding Approximate Proportional to the Integral Action to Dampen Convergence}
\label{sec_plc}
The first-order deceptive dynamics \eqref{deltadyn} are not the only dynamics of the form \eqref{deltadynamicsgeneral} that can be studied for the purpose of systematic deception. In particular, we now consider the following deceptive dynamics:
    \begin{align}\label{plcdyn}
        \left.\begin{array}{l}
        \dot{\varphi_i}=\dfrac{1}{G_{i,1}}(-\varphi_i+\varrho_i)\\ \dot{\varrho_i}=\varepsilon\varepsilon_i(J_i(x)-J_i^{\text{ref}})
        \end{array}\right\},~\delta_i=\frac{G_{i,2}}{G_{i,1}}\varrho_i-\left(\frac{G_{i,2}}{G_{i,1}}-1\right)\varphi_i,
    \end{align}
where $G_{i,2}>G_{i,1}>0$ are tunable gains. Note that the choice $G_{i,1}=G_{i,2}$ reduces \eqref{plcdyn} to the first-order dynamics \eqref{deltadyn}.  In essence, the second-order dynamics \eqref{plcdyn} incorporate a phase-lead compensation mechanism that incorporates integral action plus an approximate proportional term to improve transient performance. This improvement can be observed in the center plot of Figure \ref{figmut}, where we simulated the duopoly game with deception using \eqref{plcdyn} instead of the integral deception mechanism \eqref{deltadyn}.
%
%
The following theorem formalizes the stability properties of the deceptive NE-seeking dynamics \eqref{decgamedyn} with second-order deceptive dynamics \eqref{plcdyn}.

\vspace{0.1cm}
\begin{thm}\label{thmplc}
    Suppose that Assumption \ref{assumpw} holds, and consider the DNE seeking dynamics \eqref{decgamedyn} and \eqref{plcdyn} with $J^{\text{ref}}\in\Omega$ and $J_i$ satisfying Assumption \ref{assumpj} for all $i\in[N]$. Then there exists $\varepsilon^*$ such that for all $\varepsilon\in(0, \varepsilon^*)$ there exists $a^*$ such that for all $a\in(0, a^*)$ there exists $\omega^*$ such that for all $\omega>\omega^*$ the state $\zeta(t):=[u(t)\quad\varphi(t)\quad\varrho(t)]^\top$ converges exponentially to a $\mathcal{O}(a+\frac{1}{\omega})$ neighborhood of a point $\zeta^*$ provided $|\zeta(0)-\zeta^*|$ sufficiently small.
\end{thm}
    
\textbf{Proof:} After obtaining the average system with $\tau=\omega t$, we can apply the time scale transformation $\tau^*=\varepsilon\tau$ to obtain:
 \begin{subequations}
     \begin{align}
         \varepsilon \frac{\partial \tilde{u}}{\partial \tau^*}&=\frac{1}{\omega}\left(-k\gamma(\tilde{u},\tilde{\delta})\right)+\mathcal{O}(a)
         \\\varepsilon\frac{\partial\tilde{\varphi}}{\partial \tau^*}&=\frac{1}{\omega}\text{diag}\left(\frac{1}{G_{1,1}},...,\frac{1}{G_{n,1}}\right)(-\tilde{\varphi}+\tilde{\varrho})
         \\
         \frac{\partial\tilde{\varrho}}{\partial\tau^*}&=\frac{1}{\omega}\text{diag}\left(\varepsilon_1,...,\varepsilon_n\right)\left(J^*(\tilde{u})-J^{\text{ref}}\right)+\mathcal{O}(a),\label{deta}
     \end{align}
 \end{subequations}
where, as usual, we denote $\varphi:=[\varphi_1,...,\varphi_n]^\top$ and $\varrho:=[\varrho_1,...,\varrho_n]^\top$.
 If we disregard the $\mathcal{O}(a)$ perturbation and let $z\coloneqq [\tilde{u}\quad \tilde{\varphi}]^\top$ denote the fast state, we can obtain the quasi steady state $z^*=[g(\tilde{\varrho}),\tilde{\varrho}]^\top$ and the reduced model:
 \begin{align}
     \frac{\partial\tilde{\varrho}}{\partial\tau^*}&=\frac{1}{\omega}\text{diag}\left(\varepsilon_1,...,\varepsilon_n\right)\left(J^*(g(\tilde{\varrho}))-J^{\text{ref}}\right).\label{compred}
 \end{align}
 If $J^{\text{ref}}\in\Omega$, \eqref{compred} has an exponentially stable equilibrium point $\varrho^*\in\Delta$. Now, setting $y=z-z^*$, we obtain the boundary layer system:
 \begin{subequations}\label{compbl}
     \begin{align}
         \dfrac{\partial y_1}{\partial \tau}&=\frac{1}{\omega}\left(-k\gamma(y_1+g(\tilde{\varrho}), \Lambda^* y_2+\tilde{\varrho})\right)\label{compbl1}\\
         \dfrac{\partial y_2}{\partial \tau}&=\frac{1}{\omega}\left(-\text{diag}\left(\frac{1}{G_{1,1}},...,\frac{1}{G_{n,1}}\right)y_2\right),\label{compbl2}
     \end{align}
 \end{subequations}
\noindent
where $\Lambda^*$ is a $n\times n$ diagonal matrix with the $(i,i)$ entry being $1-G_{i,2}/G_{i,1}$.
We see that \eqref{compbl} is a cascade, and $y_2=\mathbf{0}$ is an exponentially stable equilibrium for \eqref{compbl2}. Substituting into \eqref{compbl1} gives:
\begin{equation*}
    \dfrac{\partial y_1}{\partial \tau}=\dfrac{1}{\omega}\left(-k\gamma(y_1+g(\tilde{\varrho}), \tilde{\varrho})\right),
\end{equation*}
which is exponentially stable at $y_1=0$ uniformly in $\tilde{\varrho}\in \overline{B_{R^*}(\varrho^*)}$ for some $R^*>0$. At this stage, the system reduces to the system in Theorem \ref{thmstability} but with $\zeta^*=[g(\varrho^*),\varrho^*, \varrho^*]^\top$. \QEDB 
\noindent

\vspace{0.1cm}


\section{DECEPTIVE NASH EQUILIBRIUM-SEEKING IN\\AGGREGATIVE GAMES}
\label{sec_agg}
We now study deception in games that admit non-quadratic cost functions $J_i$. Specifically, we now consider games where each player's objective function is of the following form:
\begin{equation}\label{aggcost}
    J_i(x)=c_i(x_i)+l_i(x_{-i})x_i,~~~~\forall~i\in [N],
\end{equation}
where $c_i:\mathbb{R}\to\mathbb{R}$ is twice differentiable and strongly convex, and $l_i:\mathbb{R}^{N-1}\to\mathbb{R}$ is linear. These types of games are often referred to as \emph{aggregative} games, and they are also common in the the literature of NE seeking problems in non-cooperative games, e.g., \cite{TATARENKO20203367, 7810286, 8418841}.

\vspace{0.1cm}
In aggregative games, the $i^{th}$ entry of the pseudogradient is given by
\begin{equation}
    \mathcal{G}_i(x)=\frac{\partial c_i(x_i)}{\partial x_i}+l_i(x_{-i}),
\end{equation} 
where $l_i$ can be written as
\begin{equation}
    l_i(x_{-i})=\sum_{k\neq i}\alpha_{i,k}x_k,~~~\forall~i\in[N].
\end{equation}
For convention, we will use $\alpha_{k,k}=0$ for $k\in[N]$. Under suitable conditions on $l_i$, aggregative games are strongly monotone (i.e, $\mathcal{G}$ is a strongly monotone operator). To see this, let $\kappa_i>0$ denote the strong convexity parameter of $c_i$. Setting $z:=x-y$, we obtain:
\begin{align*}
    \left(\mathcal{G}(x)-\mathcal{G}(y)\right)^\top z&=\sum_{j=1}^{N}\left(\frac{\partial c_j(x_j)}{\partial x_j}-\frac{\partial c_j(y_j)}{\partial y_j}\right)z_j+l_j(z_{-j})z_j\\
    &\ge \sum_{j=1}^{N}\left(\kappa_j z_j^2+\sum_{k\neq j}\alpha_{j,k}z_k z_j\right)\\
    &= \sum_{j=1}^{N}\kappa_j z_j^2+\sum_{k<n}(\alpha_{n,k}+\alpha_{k,n})z_k z_n\ge \kappa |z|^2,
\end{align*}
and using $ K_j=\kappa_j-\sum_{k\neq j}\frac{|\alpha_{j,k}+\alpha_{k,j}|}{2}$ it is easy to see that the aggregative game is strongly monotone with $\kappa\in(0, \min_j K_j)$ provided $K_j>0\quad\forall j$. Therefore, we make the following assumption on the game.

\vspace{0.1cm}
\begin{assumption}\label{monoassumption}
The N-player aggregative game with cost functions \eqref{aggcost} is $\kappa$-strongly monotone. \QEDB    
\end{assumption}
\vspace{-0.2cm}
\subsection{Averaging Analysis and $\delta$-Transformations of Reaction Curves}
\begin{figure}[t!]
  \centering
    \includegraphics[width=0.4\textwidth]{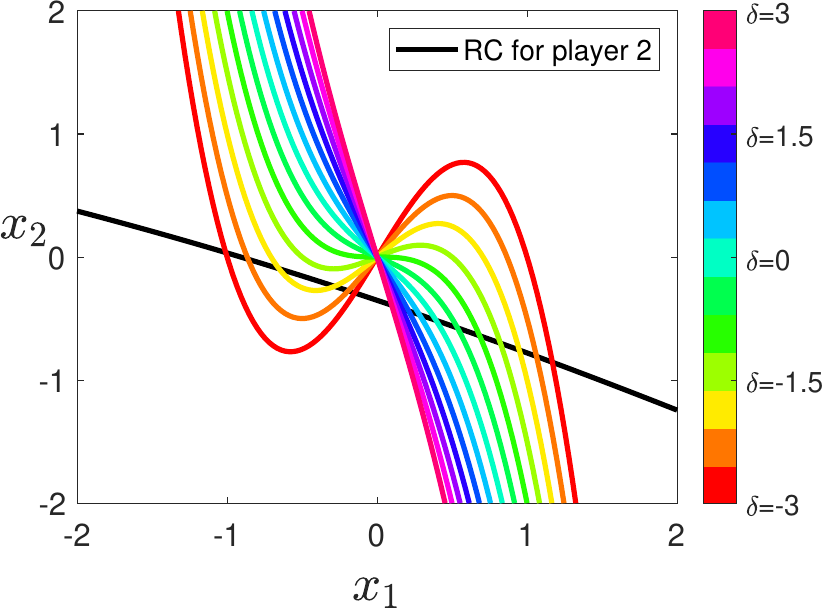}
    \caption{\small Transformations of reaction curves for a aggregative game \eqref{aggregativeexample}, where player 2 is deceptive to player 1. The colored lines represent the reactions curves of player 1 for different values of $\delta$.}\label{figagg}
    \vspace{-0.4cm}
\end{figure}

Let each player $i\in[N]$ choose its action according to the strategy \eqref{extsc0}, except for player $d$, who we assume is \emph{deceptive} and therefore implements \eqref{decx} to deceive $n_d$ players in the set $\mathcal{D}=\{d_1,...,d_{n_d}\}$. 
%
%
Using the same steps as in Section \ref{sec_nplayer}, under Assumption \ref{assumpw} the average dynamics of $u$ are given by
%
\begin{equation}\label{uavagg}
    \dot{\tilde{u}}=-k\mathcal{G}(\tilde{u})-k\delta \Psi(\tilde{u})+\mathcal{O}(a),
\end{equation}
where $\Psi(u)\in\mathbb{R}^N$ and where the $d_i^{th}$ entry of $\Psi(u)$ is now given by $\nabla_d J_{d_i}(u)=\alpha_{d_i,d}x_{d_i}$, for $i\in[n_d]$, while all the other entries are equal to $0$. As before, we first disregard the $\mathcal{O}(a)$ perturbation term in \eqref{uavagg} to proceed with our analysis. We obtain the nominal dynamics $\dot{\tilde{u}}=-k\gamma(\tilde{u}, \delta)$,
where $\gamma(\tilde{u}, \delta):=\mathcal{G}(\tilde{u})+\delta \Lambda \tilde{u}$, with $\Lambda$ being a diagonal matrix with diagonal entries equal to zero, except for the $(d_i, d_i)$ entries, which are equal to $\alpha_{d_i, d}$. Note that the change that results from deception in the average dynamics corresponds to the added term $-k\delta \Lambda \tilde{u}$. In particular, unlike the quadratic case, the effect of deception in the reaction curves of non-linear aggregative games is not any more a rotation or translation, but rather a complex nonlinear transformation that can even induce multiple critical points for the pseudogradient of the deceptive game. This behavior is illustrated in Figure \ref{figagg} for a two-player aggregative game with cost functions 
\begin{align}\label{aggregativeexample}
J_1(x)&=x_1^4+x_1^2+2x_1 x_2\\
J_2(x)&=e^{x_2}+x_2^2+1.1x_1 x_2,
\end{align}
where player 1 is deceptive to player 2.

%

\vspace{0.1cm}
\subsubsection{Preserving Monotonicity under Deception}
To study the monotonicity properties of the induced deceptive aggregative game, we first consider the case where $\delta$ is fixed. Using again $z:=x-y$, the mapping $\gamma$ satisfies:
\begin{align*}
    \left(\gamma(x, \delta)-\gamma(y, \delta)\right)^\top z&=\left(\mathcal{G}(x)-\mathcal{G}(y)\right)^\top z+\delta z^\top \Lambda z \\&\ge \sum_{j=1}^{N}K_j z_j^2+\delta\sum_{i=1}^{n_d}\alpha_{d_i, d}z_{d_i}^2\\
    &=\sum_{j\not\in\mathcal{D}}K_j z_j^2+\sum_{i\in\mathcal{D}}\left(K_i+\delta\alpha_{i,d}\right) z_i^2.
\end{align*}
Since, by Assumption \ref{monoassumption}, the nominal game is strongly monotone, to guarantee strong monotonicity of the deceptive game we just need to guarantee the condition $K_i+\delta\alpha_{i,d}>0\quad \forall i\in\mathcal{D}$. To study the set of points $\delta$ that satisfy this condition, which we now denote $\Delta$,  without loss of generality we can assume that $\alpha_{i,d}\neq 0 \quad \forall i\in\mathcal{D}$, since otherwise the $0$ cases will have no effect on such set. By direct computation we obtain the following:
\begin{lemma}\label{lemma4}
    For a $N$-player strongly monotone aggregative game where $\mathcal{D}^*=\{d\}$ and $\mathcal{D}=\{d_1,...,d_{n_d}\}$, the following estimate of $\Delta$ holds:
    \begin{align*}
    \left(\delta^-, \infty\right)&\subset\Delta\quad&&\text{if $\alpha_{i,d}>0\quad\forall i\in\mathcal{D}$}\\
    \left(-\infty, \delta^+\right)&\subset\Delta\quad&&\text{if $\alpha_{i,d}<0\quad\forall i\in\mathcal{D}$}\\
    \left(\delta^{-}, \delta^{+}\right)&\subset\Delta\quad&&\text{else},
\end{align*}
where $\delta^{-}$ and $\delta^+$ are given by
\begin{subequations}\label{deltaexpressions}
\begin{align}    
    \delta^-&=\max_{\substack{i\in\mathcal{D}\\\alpha_{i,d}>0}}\frac{-\kappa_i}{\alpha_{i,d}}+\frac{1}{\alpha_{i,d}}\sum_{k\neq i}\frac{|\alpha_{i,k}+\alpha_{k,i}|}{2}\\
    \delta^+&=\min_{\substack{i\in\mathcal{D}\\\alpha_{i,d}<0}}\frac{-\kappa_i}{\alpha_{i,d}}+\frac{1}{\alpha_{i,d}}\sum_{k\neq i}\frac{|\alpha_{i,k}+\alpha_{k,i}|}{2}.
\end{align}\vspace{-0.1cm}
\end{subequations}
\end{lemma}
The conditions of Lemma \ref{lemma4} provide some (conservative) estimates on the values $\delta$ the deceiver can use to influence the game without loosing the strong monotonicity property.

\vspace{0.1cm}
\subsubsection{When can the Deceptive Player Benefit?}
%
%
Since the main goal of deception is to obtain better profits (when possible), we now address the following question: When can a player in an aggregative game use deception to improve their own cost while maintaining closed-loop stability? 
To address this question, fix $\delta^*\in\Delta$ and pick $x^*$ such that $\gamma(x^*,\delta^*)=0$. Computing the Jacobian of $\gamma$ with respect to $x$, we obtain
\begin{align*}
    D_x\gamma(x^*, \delta^*)&=\Xi(x^*)+\delta^* \Lambda,
\end{align*}
where the matrix $\Xi$ takes the form:
\begin{equation*}
    \Xi(x^*)=\begin{bmatrix}
        c''_1(x_1^*)&\alpha_{1,2}&\alpha_{1,3}&\dots&\alpha_{1,N}\\
        \alpha_{2,1}&c''_2(x_2^*)&\alpha_{2,3}&\dots&\alpha_{2,N}\\
        \alpha_{3,1}&\alpha_{3,2}&c''_3(x^*_3)&\dots&\alpha_{3,N}\\
        \vdots&\vdots&\vdots&\vdots&\vdots\\
        \alpha_{N,1}&\alpha_{N,2}&\alpha_{N,3}&\dots& c''_N(x^*_N)
    \end{bmatrix}.
\end{equation*}
Since $\gamma(\cdot, \delta)$ is strongly monotone for all $\delta\in\Delta$, it follows that $D_x\gamma(x^*, \delta^*)$ is invertible. Then, using Lemma \ref{gqss}, we can obtain a function $g\in\mathcal{C}^1(\Delta, \mathbb{R}^N)$ which, by the implicit function theorem, also satisfies
\begin{equation}\label{dg}
    g'(\delta)=-[D_x\gamma(g(\delta), \delta)]^{-1}\Lambda g(\delta),~\quad \forall \delta\in U,
\end{equation}
where $g'$ is understood to be component wise differentiation with respect to its argument. Let $x^*$ denote the nominal NE of the averaged game dynamics (disregarding the $\mathcal{O}(a)$-perturbations) without deception, i.e when $\delta=0$, and let $\Xi^*:=\Xi(x^*)$. By inspecting \eqref{dg}, we can derive the following result, which guarantees deceptive Nash equilibrium seeking with the deceptive player $d$ always improving its payoff.
\vspace{0.1cm}
\begin{thm}\label{thmagg}
    Consider a $N-$player aggregative game with costs \eqref{aggcost} satisfying Assumption \ref{assumpj} and dynamics \eqref{decgamedyn} with $[n]:=\{d\}$, where player $d$ implements \eqref{deltadyn} on $\delta$. If $$\varepsilon_d((\Xi^*)^{-1}\Lambda x^*)_d x^*_d > 0,$$ then there is a nonempty subset $\Omega^*\subset\Omega$ such that for each $J_d^{\text{ref}}\in\Omega^*$, it follows that $J_d(u^*)<J_d(x^*)$, where $\zeta^*=[u^*\quad\delta^*]^\top$ is the point generated by Theorem \ref{thmstability}.
\end{thm}

\vspace{0.1cm}
\textbf{Proof:} We know $x^*$ satisfies:
\begin{equation*}
    \mathcal{G}_d(x^*)=c'_d(x^*_d)+l_d(x^*_{-d})=0\to l_d(x^*_{-d})=-c'_d(x^*_d).
\end{equation*}
Then, using \eqref{aggcost}, we obtain:
\begin{align}
    J_d(x^*)&=c_d(x_d^*)+l_d(x_{-d}^*)x_d^*\notag\\
    &=c_d(x_d^*)-c'_d(x^*_d)x_d^*=:\tilde{\mathcal{J}}_d(x_d^*),\label{jdeq}
\end{align}
where $\tilde{\mathcal{J}}_d:\mathbb{R}\to\mathbb{R}$. We can differentiate to extract some information about the behavior of \eqref{jdeq}:
\begin{align}    \tilde{\mathcal{J}}'_d(x_d^*)&=-c''_d(x_d^*)x_d^*\label{jaggd}
\end{align}
which vanishes at $x_d^*=0$, is positive for $x_d^*<0$ and negative for $x_d^*>0$  by the strong convexity of $c_d$. Hence $\tilde{\mathcal{J}}_d(x_d^*)$ achieves a global maximum at $x_d^*=0$ and decreases without bound as $|x_d^*|\to\infty$. In particular, this tells us that no matter the value of $x^*_d$, there is a ``direction" such that $J_d$ decreases when $x_d^*$ is moved in that ``direction". Since player $d$ seeks to minimize $J_d$, it is most desirable to find $\delta\in\Delta$ such that $|x_d^*|$ is maximized.
As we have demonstrated above, we can use the implicit function theorem to obtain $g\in\mathcal{C}^{1}(\Delta,\mathbb{R}^N)$ such that $\tilde{u}=g(\delta)$ is the DNE of the unperturbed \eqref{uavagg} for $\delta\in \Delta$ and $g'(0)=-(\Xi^*)^{-1}\Lambda x^*$. Since $\varepsilon_d((\Xi^*)^{-1}\Lambda x^*)_d x^*_d > 0$, we know $g_d'(0)\neq 0$, so by observing \eqref{jdeq} and \eqref{jaggd} we know there is some open interval $E_1\subset \Delta$ such that $J_d(g(\delta))<J_d(x^*)\quad\forall \delta\in E_1$, where $E_1$ is either of the form $(-R,0)$ or $(0, R)$ for some $R>0$. Furthermore, by continuity we can also find an open neighborhood $E_2$ of 0 such that $ g_d'(\delta) g_d(\delta)>0,\quad\forall \delta\in E_2$. Let $E:=E_1\cap E_2$, which is also an open interval. Then we set $\Omega^*:=\left(\inf_{\delta\in E} J_d(g(\delta)), J_d(x^*)\right)$,
which completes the proof. \hfill $\blacksquare$
%
%
\begin{figure}[t!]
  \centering
    \includegraphics[width=0.4\textwidth]{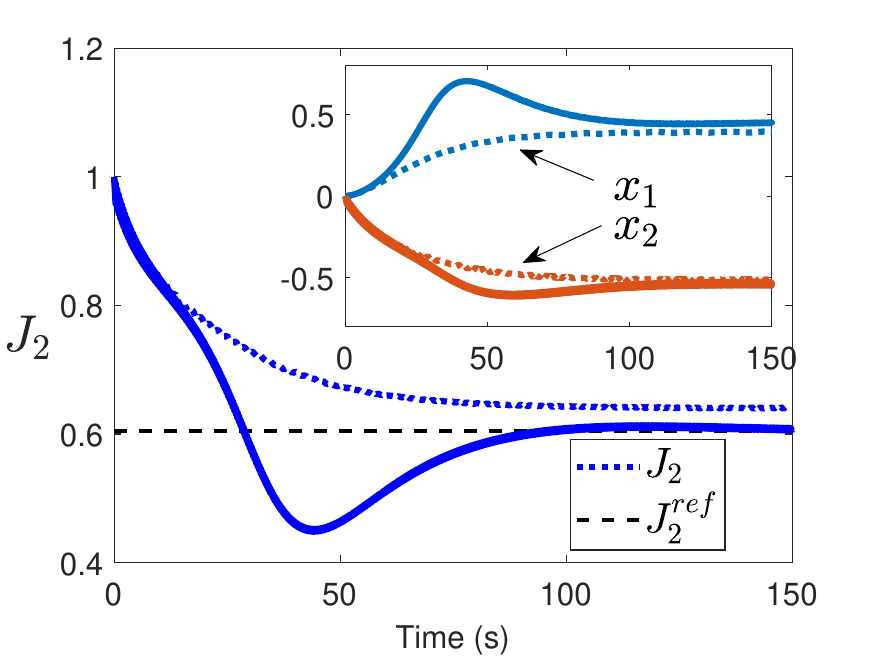}
    \caption{\small Action and price convergence for an aggregative game. Here we use $a=0.01, k=0.03, \omega_1=470.75, \omega_2=330$. The solid lines represent the actions and cost in the deceptive game.}\label{aggexx}
    \vspace{-0.3cm}
\end{figure}

From the proof of Theorem \ref{thmagg}, the expression in \eqref{jaggd} can be used to establish the following Corollary for two player games:
\begin{cor}
    Consider a two player strongly monotone aggregative game where player 1 is deceptive to player 2. Let $(\underline{\delta}, \overline{\delta})\subset\Delta$ such that $0\in(\underline{\delta}, \overline{\delta})$, where $\underline{\delta}, \overline{\delta}$ are given by \eqref{deltaexpressions}. Furthermore, assume $x_2^*\neq 0$. Then, at least one of the following holds
    \begin{enumerate}[(a)]
        \item $J_1(g(\delta))$ is strictly decreasing on $(0, \overline{\delta})$
        \item $J_1(g(\delta))$ is strictly increasing on $(\underline{\delta}, 0)$
    \end{enumerate}
    where $g(\delta)=x_\delta$ is obtained from Lemma \ref{gqss}. \QEDB 
\end{cor}

\textbf{Proof:}
    In this setting, we have:
    \begin{equation}
    \gamma(x,\delta)=\begin{bmatrix}
        c'_1(x_1)+\alpha_{1,2}x_2\\
        c'_2(x_2)+\alpha_{2,1}x_1+\delta\alpha_{2,1}x_2
    \end{bmatrix}.\label{gtild2}
\end{equation}
It follows that $g_2(\delta)$ is injective. Indeed, suppose there exists $\delta_a, \delta_b\in\Delta$ with $\delta_a\neq \delta_b$ and $g_2(\delta_a)=g_2(\delta_b)$. Then the top entry of \eqref{gtild2} implies $g(\delta_a)=g(\delta_b)$, and the bottom entry gives $(\delta_a-\delta_b)\alpha_{2,1}g_2(\delta_a)=0$, which implies $g_2(\delta_a)=0$. But this means $g(\delta_a)\neq x^*$ is also  a NE of the non-deceptive game, which contradicts the assumption that the NE is unique. Then $g_2(\delta)$ is strictly monotone on $\Delta$, and since the top entry of \eqref{gtild2} is a strictly monotone curve in the $\mathbb{R}^2$ plane we also conclude $g_1(\delta)$ is strictly monotone on $\Delta$. By combining this with the observation from \eqref{jaggd} (but replace $x^*$ with $g(\delta)$), we obtain the result. \hfill $\blacksquare$

This result suggests that one viable strategy for the deceiver in a 2-player aggregative game is to tune $\delta$ monotonically. In particular, if the conditions of Theorem \ref{thmagg} are satisfied and the deceiver is able to improve their cost by perturbing $\delta$ in some ``direction" away from 0, then they can continue tuning $\delta\in \Delta$ in that same ``direction". 

\vspace{-0.2cm}
\subsection{Numerical Example}
Consider a two-player aggregative game with cost given by \eqref{aggregativeexample}.
%
%
It is straightforward to verify that $\kappa_1=\kappa_2=2, \alpha_{1,2}=2$ and $\alpha_{2,1}=1.1$. It can also be verified that this game is strongly monotone since $\kappa_j-\frac{|\alpha_{2,1}+\alpha_{1,2}|}{2}=2-\frac{|3.1|}{2}>0$. Suppose player 2 is deceptive to player 1. Then we can compute $\mathcal{G}$ and $\gamma$ as follows:

\vspace{-0.3cm}
\begin{small}
\begin{align*}
    \mathcal{G}=\begin{bmatrix}
        4x_1^3+2x_1+2x_2\\
        e^{x_2}+2x_2+1.1x_1
    \end{bmatrix},\gamma=\begin{bmatrix}
        4x_1^3+2x_1+2x_2+2\delta x_1\\
        e^{x_2}+2x_2+1.1x_1
    \end{bmatrix},
\end{align*}
\end{small}
and we also have that $\Delta=(-0.225, \infty)$. Note that setting the second component of $\gamma$ equal to 0 results in a monotonic curve in the $\mathbb{R}^2$ plane. That is, as $x_2$ increases we must have that $x_1$ decreases. This relationship would of course be reversed if $\alpha_{2,1}$ were negative. Figure \ref{figagg} presents a visualization for how deception affects the reaction curve for player 1 in this setting. An interesting observation is that, similar to quadratic games, all the reaction curves of the oblivious player intersect at a single point. However, in the aggregative game considered in this example, the ``transformation" induced by dynamic deception on the reaction curves is not anymore a rotation. The shared point in this example is the point on player 1's reaction curve where $x_1=0$, which is $\mathbf{0}$. In this case, player 2 sets $J_2^{\text{ref}}=0.605$, achieved at $\delta=-0.22\in\Delta$. The convergence of $x_i$ and $J_2$ are visualized in Figure \ref{aggexx}. Additional analytical and numerical examples can be found in the supplemental material.
\section{CONCLUSION}
\label{ref_conclusions}
We introduced the problem of model-free Nash equilibrium-seeking with deception for non-cooperative games with finitely many players. The setting considered incorporates oblivious and deceiving players, and studies the \emph{stable} behaviors that emerge under a class of model-free (or payoff-based) algorithms that rely on simultaneous exploration and exploitation policies implemented in games with \emph{asymmetric information}. The geometric and structural properties of the induced deceptive games and deceptive Nash equilibria were studied, as well as the stability properties of the dynamics under benevolent deception, mutual deception, and using high-order deceptive dynamics. 
Our results open the door to several new research questions at the intersection of adaptive and learning systems, and deception in game theory.

\vspace{-0.3cm}
\bibliographystyle{IEEEtran}
\bibliography{Biblio.bib,Biblio2.bib,dissertation.bib}

\vspace{-0.8cm}
 \begin{IEEEbiography}[{\includegraphics[width=1in,height=1.25in,clip,keepaspectratio]{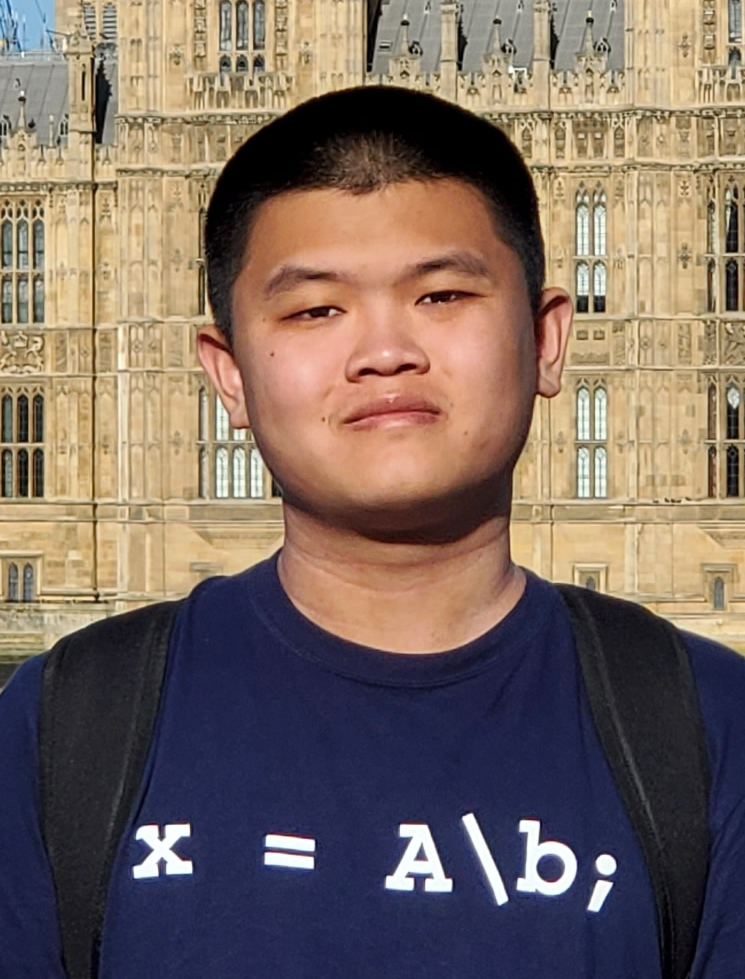}}]{Michael Tang} 
 is a Ph.D. student in the Department of Electrical and Computer Engineering, at the University of California, San Diego. He received his B.S. degrees in Electrical Engineering and Applied Mathematics from UCSD in 2023. His research interests are in control theory, optimization, game theory, and dynamical systems. He was a finalist for the Best Student Paper award at the 2024 American Control Conference.
 \end{IEEEbiography}

\vspace{-0.8cm}
  \begin{IEEEbiography}[{\includegraphics[width=1in,height=1.25in,clip,keepaspectratio]{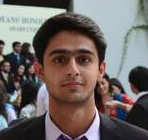}}]{Umar Javed} 
received in 2017 his B.S.
degree in Elect. Eng. with minors in
Computer Science (AI/ML) and Psychology from the Lahore University of
Management Sciences, Pakistan. He
also received his MS degree in 2019,
and his Ph.D. degree in 2023, both in Electrical and  Computer Engineering from
the University of Colorado, Boulder. He is currently affiliated with Amazon, in Seattle, WA.
 \end{IEEEbiography}

\vspace{-0.8cm}
\begin{IEEEbiography}[{\includegraphics[width=1in,height=1.25in,clip,keepaspectratio]{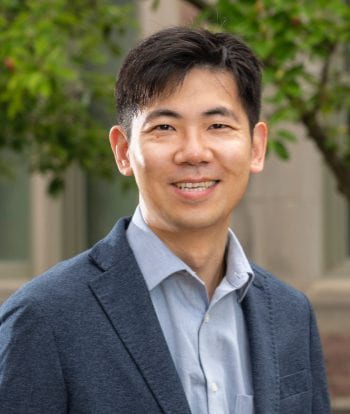}}]{Xudong Chen} 
is Associate Professor in the Department of Electrical \& Systems Engineering at Washington University in St. Louis. Prior to that, he was an assistant professor in the Department of Electrical, Computer, and Energy Engineering at the University of Colorado, Boulder. Chen earned a BS degree in electronics engineering from Tsinghua University, China, in 2009, and a PhD degree in electrical engineering from Harvard University in 2014. Chen received the 2023 A.V. ‘Bal’ Balakrishnan Early Career Award, the 2021 Donald P. Eckman Award, an NSF CAREER Award in 2021, and the AFOSR Young Investigator Award in 2020.
 \end{IEEEbiography}

\vspace{-0.8cm}
 \begin{IEEEbiography}[{\includegraphics[width=1in,height=1.25in,clip,keepaspectratio]{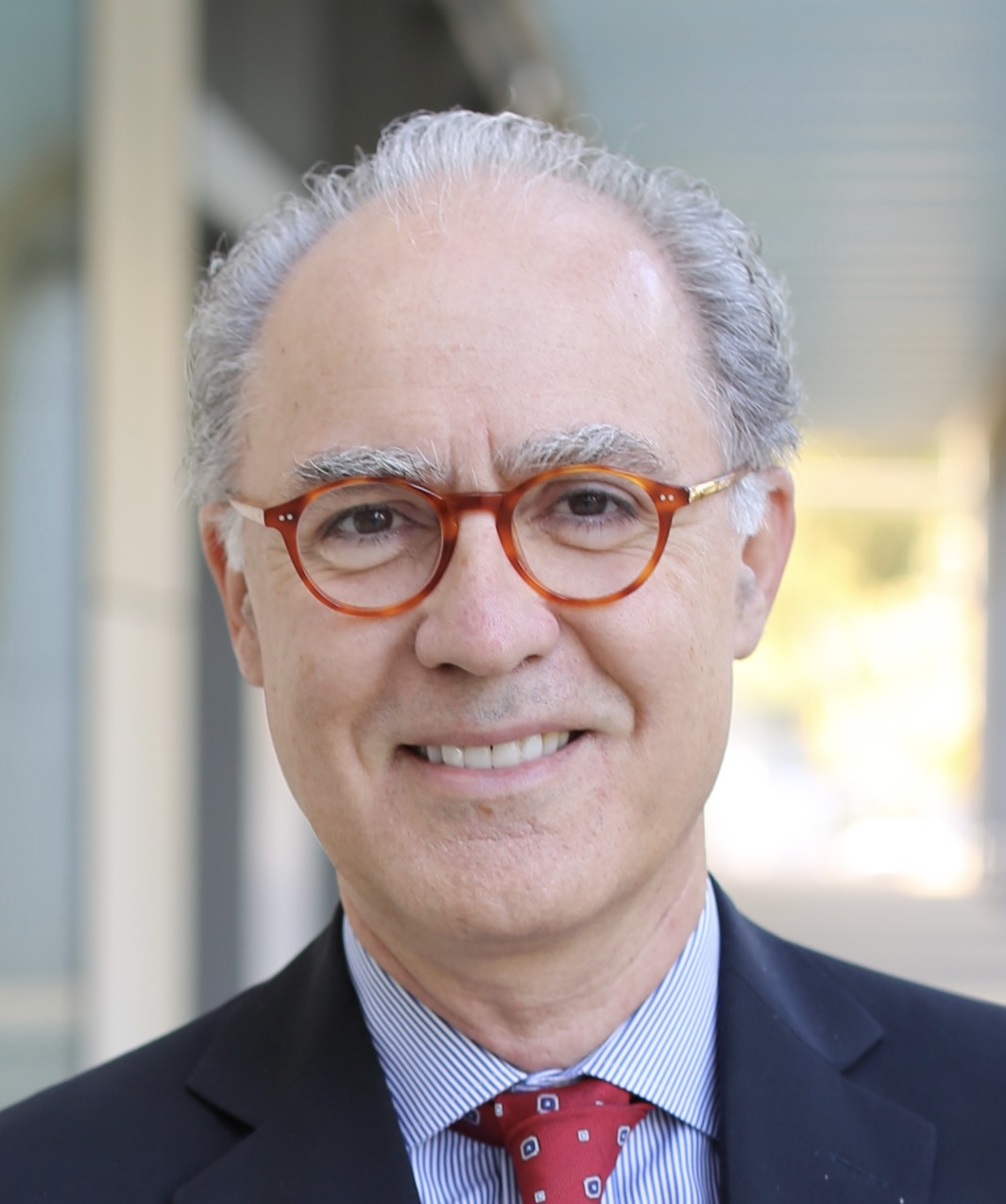}}]{Miroslav Krsti\'c} is Distinguished Professor at 
 UC San Diego. He is Fellow of IEEE, IFAC, ASME, SIAM, AAAS, IET, AIAA (Assoc. Fellow), and  Serbian Academy of Sciences and Arts. He has received the Bellman Award, Bode Lecture Prize,  Reid Prize,  Oldenburger Medal, Nyquist Lecture Prize, Paynter  Award, Ragazzini  Award, IFAC Nonlinear Control Systems Award,  Distributed Parameter Systems Award,  Adaptive and Learning Systems Award, Chestnut Award, AV Balakrishnan Award, CSS Distinguished Member Award, the PECASE, NSF Career, and ONR YIP, and the Schuck and Axelby paper prizes. He serves as Editor-in-Chief of Systems $\&$ Control Letters and Senior Editor in Automatica. 
 \end{IEEEbiography}
 
\vspace{-0.7cm}
 \begin{IEEEbiography}[{\includegraphics[width=1in,height=1.25in,clip,keepaspectratio]{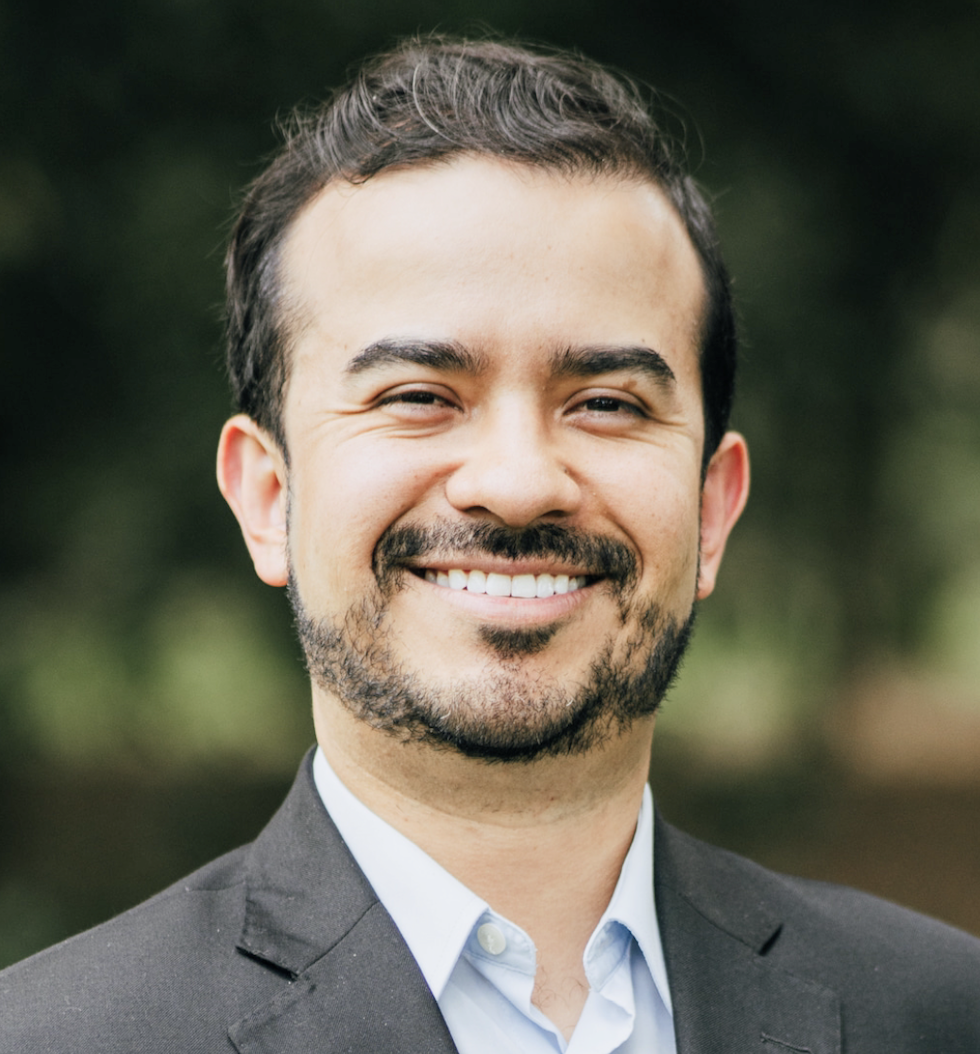}}]{Jorge I. Poveda} 
 is Assistant Professor in the Electrical and Computer Engineering Department at UC San Diego. Prior to that,
he was Assistant Professor in the Department
of Electrical, Computer, and Energy Engineering
at the University of Colorado, Boulder, and a Postdoctoral scholar at Harvard. He received 
 his M.Sc. (2016) and Ph.D. (2018) degrees in Electrical and Computer Engineering from UC Santa Barbara. 
 He has received the NSF CRII and CAREER awards, the AFOSR and SHPE Young Investigator Award, the UCSB-CCDC Outstanding Scholar Fellowship  and Best Ph.D. Thesis award, the 2023 AACC Donald P. Eckman award, and the 2025 UC San Diego Engineering Early Faculty Development Award. 
 \end{IEEEbiography}

\section*{Appendix: Supplemental Material}
\subsection{Local Practical Exponential Stability of System \eqref{thm1spf}}
We consider the dynamics \eqref{thm1spf}, without the $\mathcal{O}(a)$-perturbation, which have the form:
\begin{subequations}\label{sys}
\begin{align}
    \varepsilon \dot{u}&=-k\gamma(u, \delta)\label{fast}\\
    \dot{\delta}&=\text{diag}(\varepsilon_1,...,\varepsilon_n)(J^*(u)-J^{\text{ref}}),\label{slow}
\end{align}
\end{subequations} 
where ${J}^*(\tilde{u}):=[J_1(\tilde{u}),...,J_n(\tilde{u})]^\top$, and where $J^{\text{ref}}$ is assumed to be attainable. Let $u=g(\delta)$ be the quasi-steady state of \eqref{fast}, which, by Lemma \ref{gqss}, is well-defined. By substituting in \eqref{slow}, we obtain the reduced dynamics 
\begin{equation}
        \dot{\delta}=\text{diag}(\varepsilon_1,...,\varepsilon_n)(J^*(g(\delta))-J^{\text{ref}}),\label{red}
\end{equation}
which is a $\mathcal{C}^1$ vector field given that $J^*$ is $\mathcal{C}^2$ and $g\in\mathcal{C}^1$. Since $J^{\text{ref}}\in\Omega$, system \eqref{red} has a locally exponentially stable equilibrium point $\delta^*$. Therefore, by a standard converse Lyapunov theorem \cite[Thm. 4.14]{khalil}, there exists $r>0$ and a Lyapunov function $V$ satisfying
\begin{subequations}
       \begin{align}
        c_1 |\delta-\delta^*|^2\le V(\delta)&\le c_2|\delta-\delta^*|^2\label{vbd1}\\
        \dfrac{\partial V}{\partial \delta}\text{diag}(\varepsilon_1,...,\varepsilon_n)(J^*(g(\delta))-J^{\text{ref}})&\le -c_3 |\delta-\delta^*|^2\\
        \left|\dfrac{\partial V}{\partial \delta}\right|&\le c_4 |\delta-\delta^*|,
    \end{align} 
\end{subequations}
for all $\delta\in B_r(\delta^*)$. 

Similarly, by defining $y=u-g(\delta)$, we can obtain the boundary layer dynamics of system \eqref{sys}, which are given by
\begin{equation}\label{bl}
        \dot{y}=-k\gamma(y+g(\delta), \delta).
\end{equation}
Since by Lemma \ref{gqss} the set $\Delta$ is a non-empty open set, there exists $r^*\in (0, r)$ such that $\overline{B_{r^*}(\delta^*)}\subset \Delta$. Let $K:=\overline{B_{r^*}(\delta^*)}$ which is compact, and let $\Gamma(y, \delta):=\gamma(y+g(\delta), \delta)$. Then, since $\frac{\partial \Gamma}{\partial y}(0, \delta)=\frac{\partial \gamma}{\partial u}(g(\delta), \delta)$ for $\delta\in K\subset\Delta$, we have that:
\begin{equation*}
    \text{Re}\left(\lambda\left(\dfrac{\partial \Gamma}{\partial y}(0, \delta)\right)\right) <0\quad \forall \delta\in K .
\end{equation*} 
But since $K$ is compact, 
\begin{equation}
        \max_{\delta\in K}\left(\text{Re}\left(\lambda\left(\dfrac{\partial \Gamma}{\partial y}(0, \delta)\right)\right)\right)\le c<0.
\end{equation}
Thus, by \cite[Exercise 11.5]{khalil} the origin of \eqref{bl} is exponentially stable uniformly in $\delta\in K$, and by \cite[Lemma 9.8]{khalil} there exists $\rho>0$ and a Lyapunov function $W(y, \delta)$ satisfying
\begin{subequations}
        \begin{align}
            b_1 |y|^2\le W(y, \delta)&\le b_2 |y|^2\\
            -k\dfrac{\partial W}{\partial y}\gamma(y+g(\delta), \delta)&\le -b_3 |y|^2\label{wbd1}\\
            \left|\dfrac{\partial W}{\partial y}\right|\le b_4 |y| \quad  \left|\dfrac{\partial W}{\partial \delta}\right|&\le b_5 |y|^2.        
            \end{align}
\end{subequations}
for $y\in B_\rho (0)$.
Using again the change of variables $y=u-g(\delta)$, we can write the original dynamics \eqref{sys} as follows:
\begin{subequations}\label{newsys}
            \begin{align}
                \varepsilon \dot{y}&=-k\gamma(y+g(\delta), \delta)\notag\\
                &~~~~~~~~~-\varepsilon \dfrac{\partial g}{\partial \delta} \text{diag}(\varepsilon_1,...,\varepsilon_n)(J^*(y+g(\delta))-J^{\text{ref}})\\
                \dot{\delta}&=\text{diag}(\varepsilon_1,...,\varepsilon_n)(J^*(y+g(\delta))-J^{\text{ref}}).
            \end{align}
\end{subequations}
To study this system, we consider the Lyapunov function candidate $\Psi=V+W$, whose time derivatives along the trajectories of \eqref{newsys} satisfies

\vspace{-0.2cm}
\begin{small}
\begin{align*}
            \dot{\Psi}&=\left(\dfrac{\partial V}{\partial \delta}+\dfrac{\partial W}{\partial \delta}-k\dfrac{\partial W}{\partial y}\dfrac{\partial g}{\partial \delta}\right)\text{diag}(\varepsilon_1,...,\varepsilon_n)\Bigg(J^*(y+g(\delta))\\
            &~~~~~-J^*(g(\delta))+J^*(g(\delta))-J^{\text{ref}}\Bigg)-\dfrac{k}{\varepsilon}\dfrac{\partial W}{\partial y}\gamma(y+g(\delta), \delta).
\end{align*}
\end{small}
Note that by continuity and local Lipschitz continuity the functions $J^*(y+g(\delta))-J^*(g(\delta))$, $J^*(g(\delta))-J^{\text{ref}}$ and all of their partial derivatives are bounded in $(y, \delta)\in B_\rho(0)\times K$ and satisfy the following estimates in such set:
\begin{subequations}
            \begin{align}
                |\text{diag}(\varepsilon_1,...,\varepsilon_n)\left(J^*(y+g(\delta))-J^*(g(\delta))\right)|&\le L_1 |y|\\
                |\text{diag}(\varepsilon_1,...,\varepsilon_n)\left(J^*(g(\delta))-J^{\text{ref}}\right)|&\le L_2 |\delta-\delta^*|\\
                \left|\dfrac{\partial g}{\partial \delta}\right|&\le \kappa.\label{est3}
            \end{align}
\end{subequations}
It follows that, for all $y\in B_\rho(0)$, we have:
\begin{align*}
            \dot{\Psi}&\le -c_3 |\delta-\delta^*|^2+c_4 L_1 |\delta-\delta^*||y|+b_5 L_1 |y|^3\\
            &~~~~+b_5 L_2 |\delta-\delta^*||y|^2+k\kappa b_4 L_1 |y|^2+k\kappa b_4 L_2 |y||\delta-\delta^*|\\
            &~~~~-\dfrac{b_3}{\varepsilon}|y|^2\\
            &\le -c_3 |\delta-\delta^*|^2+c_4 L_1 |\delta-\delta^*||y|+b_5 L_1\rho |y|^2\\
            &~~~~+b_5 L_2\rho |\delta-\delta^*||y|+k\kappa b_4 L_1 |y|^2+k\kappa b_4 L_2 |y||\delta-\delta^*|\\
            &~~~~-\dfrac{b_3}{\varepsilon}|y|^2,\\
            &=-\begin{bmatrix}
                |\delta-\delta^*|\\ |y|
    \end{bmatrix}^\top\Lambda_\varepsilon\begin{bmatrix}
                |\delta-\delta^*|\\ |y|
            \end{bmatrix},
\end{align*}
where, using $c^*:=c_4 L_1 +b_5 L_2\rho+k\kappa b_4 L_2$ and $d:=k\kappa b_4 L_1+b_5 L_1 \rho$, we have
\begin{equation*}
            \Lambda_\varepsilon=\begin{bmatrix}
                c_3 &-\frac12 c^*\\-\frac12 c^*& \dfrac{b_3}{\varepsilon}-d
            \end{bmatrix}.
\end{equation*}
Pick $\varepsilon^*$ such that $\Lambda_\varepsilon\succ 0$ for all $\varepsilon\in (0, \varepsilon^*)$. Such $\varepsilon^*$ always exists by \cite[pp. 452]{khalil}.  For these values of $\varepsilon$, we have
\begin{align*}
            \dot{\Psi}&\le -\underline{\lambda}(\Lambda_\varepsilon)\left(|\delta-\delta^*|^2+|y|^2\right)\le \frac{-\underline{\lambda}(\Lambda_\varepsilon)}{\max(b_2, c_2)}\Psi.
\end{align*}
If we let $\vartheta=\frac{\underline{\lambda}(\Lambda_\varepsilon)}{\max(b_2, c_2)}$,
we have
\begin{equation*}
            \Psi(\delta(t), y(t))\le e^{-\vartheta t}\Psi(\delta(0), y(0))
\end{equation*}
Combining this with \eqref{vbd1} and \eqref{wbd1} yields
\begin{figure*}[t!]
  \centering
  \includegraphics[width=0.42\textwidth]{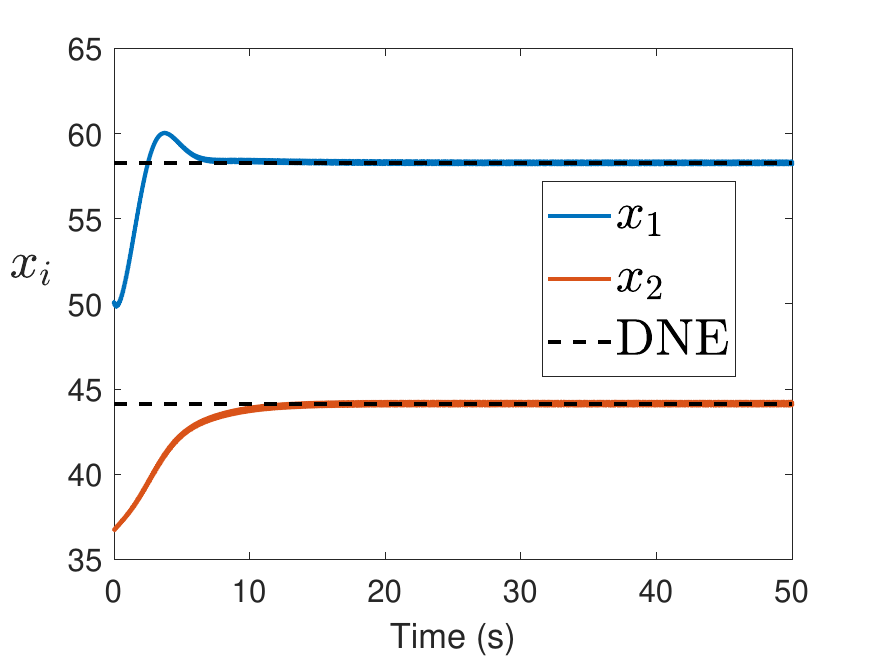}
    \includegraphics[width=0.42\textwidth]{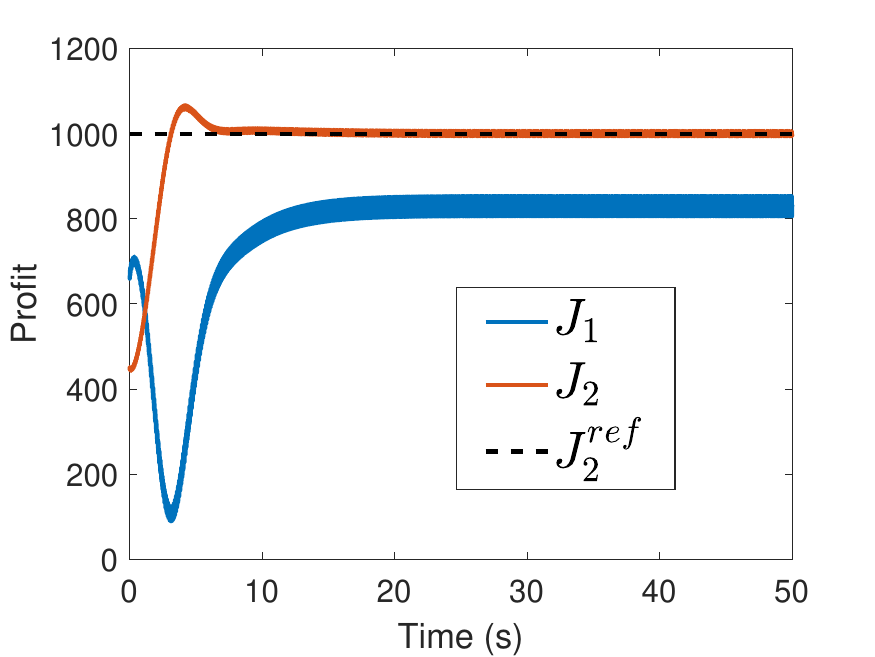}
    \caption{\small Profit and price convergence for the duopoly when players implement the compensator dynamics \eqref{plcdyn} with gains $G_{2,1}=0.5, G_{2,2}=3.9$.}\label{duocompp}
    \vspace{-0.2cm}
\end{figure*}
\begin{equation*}
            \left|\begin{bmatrix}
                \delta(t)-\delta^*\\ y(t)
            \end{bmatrix}\right|\le e^{-\frac12\vartheta t}\sqrt{\dfrac{\max(b_2, c_2)}{\min(b_1, c_1)}}\left|\begin{bmatrix}
                \delta(0)-\delta^*\\ y(0)
            \end{bmatrix}\right|
\end{equation*}
But we also have $|u(t)-u^*|\le |y(t)|+|g(\delta)-g(\delta^*)|\le |y(t)|+\kappa |\delta-\delta^*|$ where we use \eqref{est3} and the fact that $u^*=g(\delta^*)$. Thus, we obtain
\begin{align*}
            \left|\begin{bmatrix}
                \delta(t)-\delta^*\\ u(t)-u^*
            \end{bmatrix}\right|\le \max(\sqrt{1+2\kappa^2}, \sqrt{2})\left|\begin{bmatrix}
                \delta(t)-\delta^*\\ y(t)
            \end{bmatrix}\right|.
\end{align*}
Combining the previous two bounds, we obtain the local exponential stability result for the nominal system \eqref{sys}. Local practical exponential stability as $a\to0^+$ follows now directly by standard robustness results for perturbed smooth systems \cite[Thm. 17]{Goebel_CSM}. \hfill $\blacksquare$
\subsection{Duopoly with Deception: One-Time scale Analysis and the Effect of Greedy Players}
\subsubsection{Payoff Reference}
In this section we directly assess the stability of \eqref{duoavg} interconnected with \eqref{duodelta} without studying the reduced and boundary-layer dynamics. Linearizing around an equilibrium $(x^*, \delta^*)$ yields the 3 dimensional system:
\begin{align}\label{duo_lin}
    \dot{\zeta}=\begin{bmatrix}
        -\mathcal{Q_{\delta^*}}& -\overline{\mathcal{Q}}x^*-\overline{B}\\
        \varepsilon ({x^*}^\top Q+b^\top)& 0
    \end{bmatrix}\zeta:=A(x^*, \delta^*)\zeta
\end{align}
where, $\delta^*\in\Delta$ and, as we recall, $\mathcal{Q}_\delta=\mathcal{Q}+\delta\overline{\mathcal{Q}}$. Let $P_{\mathcal{Q}}(s)$ denote the characteristic polynomial of $-\mathcal{Q}_{\delta^*}$. Furthermore, if we let $P_A(s)$ denote the characteristic polynomial of $A(x^*, \delta^*)$, it can be verified by direct computation that there exists some degree 1 polynomial $p(s)$ such that:
\begin{align*}
    P_A(s)&=sP_{\mathcal{Q}}(s)+\varepsilon p(s) 
\end{align*}
If we denote $P_{\mathcal{Q}}(s)=s^2+a_1 s+a_0$ and $p(s)=a_1^* s+a_0^*$, we obtain:
\begin{equation*}
    P_A(s)=s^3+a_1 s^2+(a_0+\varepsilon a_1^*)s+\varepsilon a_0^*
\end{equation*}
Since $\delta^*\in\Delta$ implies $-\mathcal{Q}_{\delta^*}$ is Hurwitz, we know $a_1, a_0>0$. We can first fix $\text{sgn}(\varepsilon)$ to satisfy $\varepsilon a_0^*>0$. Then, for $A(x^*, \delta^*)$ to be Hurwitz, the Routh-Hurwitz criterion implies that we need $a_0+\varepsilon a_1^*>0$ and $a_1 (a_0+\varepsilon a_1^*)-\varepsilon a_0^*>0$. Fortunately, since $a_1, a_0>0$, we can find $\varepsilon^*$ such that these conditions are satisfied whenever $|\varepsilon|<\varepsilon^*$. One possible choice of $\varepsilon^*$ is
\begin{equation*}
    \varepsilon^*=\min\left(\frac{a_0}{|a_1^*|}, \frac{a_1 a_0}{|a_1 a_1^*-a_0^*|}\right).
\end{equation*} 
Note that here we establish local stability of the deceptive duopoly by linearization, even though for the main result we use singular perturbation theory. If we set $m_1=m_2$, \eqref{duo_lin} can be greatly simplified into the following form:
\begin{equation*}
    \dot{\zeta}=\begin{bmatrix}
        -\dfrac{1}{2p}-\dfrac{S_d}{2\sqrt{|J_2^{\text{ref}}p|}} & \dfrac{1}{p} & 2\sqrt{\dfrac{|J_2^{\text{ref}}|}{p}}\\
        \dfrac{1}{p} & -\dfrac{2}{p} & 0\\
        -\varepsilon \sqrt{\dfrac{|J_2^{\text{ref}}|}{p}} & 0 & 0
    \end{bmatrix}\zeta
\end{equation*}
where the matrix has the following characteristic polynomial:
\begin{align*}
    P_A(s)&=s^3+\left(\frac{5}{2p}+\dfrac{S_d}{2\sqrt{|J_2^{\text{ref}}p|}}\right)s^2\\&~~~+\left(2\varepsilon\dfrac{|J_2^{\text{ref}}|}{p}+\dfrac{S_d}{p\sqrt{|J_2^{\text{ref}}p|}}\right)s+4\varepsilon\dfrac{|J_2^{\text{ref}}|}{p^2}
\end{align*}
which turns out to lead to a Hurwitz matrix for all $\varepsilon>0$. The previous approach reveals some insight about deception on duopolies with equal marginal costs:
\begin{enumerate}
\item As the deceiver becomes more greedy (i.e., $J_2^{\text{ref}}$ grows, the effect of the demand $S_d$ on the stability properties of the algorithm vanishes.
\item The rate at which the quantity $p$ (the market's preference for player 1) increases affects the terms in the polynomial at different rates. However, for any $p>0$, a greedy player 2 can always attain a desired profit $J_2^{\text{ref}}$.
\item Larger values of $\varepsilon$ can impact only the last two coefficients of the polynomial, indicating that no matter how impatient is the deceiver, fundamental limits imposed by the game would dictate the final transient and rate of convergence to the DNE.
\end{enumerate}
Therefore, the previous observations indicate that a deceiver that is greedy for very high profit can attain any such profit, and pursue this with arbitrarily high integral gain (impatiently), irrespective of the market preference for the victim.

This approach of directly linearizing the interconnection reveals opportunities to relax conditions on the gain of the integrator of the deceiving agent in different types of games, but it presents several challenges when generalizing to higher dimensions and would require defining a new notion of ``attainability". However, the idea remains worth exploring for future work.
\subsubsection{Price Reference}
While the results in Sections \ref{probl_statement}-\ref{sec_agg} focused on deception dynamics that seek to attain a desired profit $J_i^{\text{ref}}$, it is possible to consider other types of deception dynamics. For instance, consider the duopoly in Example \ref{example1}, and suppose player 2 instead wants to use deception to stabilize a desired price $x_2^{\text{ref}}$. Then, player 2 can implement the following integral update law on $\delta$:
\begin{equation*}
    \dot{\delta}=\varepsilon (u_2-u_2^{\text{ref}}).
\end{equation*}
We can again linearize the dynamics to obtain
\begin{align*}
    \dot{\zeta}=\begin{bmatrix}
        -\mathcal{Q_{\delta^*}}& -\overline{\mathcal{Q}}x^*-\overline{B}\\
        0\quad \varepsilon& 0
    \end{bmatrix}\zeta:=A(x^*, \delta^*)\zeta.
\end{align*}
This reduces to the same problem studied in \eqref{duo_lin}, but now we have a simpler expression for $a_0^*:$
\begin{equation*}
    a_0^*=(-\mathcal{Q}_{\delta^*})_{2,1}(-\overline{\mathcal{Q}}x^*-\overline{B})_1
\end{equation*}
If $a_0^*>0$, player 2 should set $\varepsilon>0$, otherwise they should set $\varepsilon<0$. Then, as done previously, we could compute the characteristic polynomial and verify with the Routh Hurwitz criterion that this interconnection is locally exponentially stable for $|\varepsilon|$ small, thus deceiving player 1 into using a price $u_1$ for which the pair $(u_1,u_2^{\text{ref}})$ is a DNE.
\subsection{Additional Numerical Examples}
To illustrate how the deception dynamics with compensator \eqref{plcdyn} can reduce the oscillations in the system, we simulate the same duopoly game considered in Example 1 and Figure 2. The results are presented in Figure \ref{duocompp}, and they showcase how system \eqref{plcdyn} can induce stable deception without oscillations in the response such as those shown in Figure 2.

\end{document}